\documentclass[acmsmall]{acmart}

\newcommand{\harun}[1]{\textcolor{black}{#1}}

\AtBeginDocument{%
  \providecommand\BibTeX{{%
    \normalfont B\kern-0.5em{\scshape i\kern-0.25em b}\kern-0.8em\TeX}}}
\usepackage{tabularx}
\usepackage{tikz}
\usetikzlibrary{calc}
\tikzset{circle mark/.pic={
  \draw [path picture={%
    \fill (path picture bounding box.south west) |-  
      ($(path picture bounding box.south east)!#1/100!(path picture bounding box.north east)$) 
      |- cycle;}] circle [radius=1ex];
}}

\usepackage{rotating}
\usepackage[nobiblatex]{xurl}
\usepackage{tablefootnote}
\usepackage{blindtext}
\usepackage{adjustbox}
\usepackage{graphicx}
\usepackage[para]{threeparttable}
\usepackage{array}
\usepackage{multirow}
\usepackage{enumitem}
\usepackage{microtype}
\usepackage{float}
\usepackage{booktabs,tabularx}
\usepackage{xcolor}
\usepackage{colortbl}
\usepackage{lipsum}
\usepackage{supertabular}
\newcolumntype{L}[1]{>{\raggedright\arraybackslash}p{#1}}
\usepackage{titlesec}
\usepackage[hang]{subfigure}
\usepackage{hyperref, longtable}
\PassOptionsToPackage{hyphens}{url}
\usepackage{hyperref}
\usepackage{xurl}
\usepackage{makecell}
\newcommand{\ahmet}[1]{{\textcolor{black}{#1}}}

\acmJournal{CSUR}

\begin{document}
\sloppy

\title{A Survey on Ransomware: Evolution, Taxonomy, and Defense Solutions}


\author{Harun Oz}
\email{hoz001@fiu.edu}
\author{Ahmet Aris}
\email{aaris@fiu.edu}
\author{Albert Levi}
\email{levi@sabanciuniv.edu}
\author{A. Selcuk Uluagac}
\email{suluagac@fiu.edu}
\affiliation{%
  \institution{\\Cyber-Physical Systems Security Lab, Florida International University, Miami, Florida, USA}
  \postcode{33174}
}
\affiliation{%
  \institution{\\Faculty of Engineering and Natural Sciences, Sabancı University, Istanbul, Turkey}
}

\renewcommand{\shortauthors}{Oz, et al.}

\begin{abstract}
In recent years, ransomware has been one of the most notorious malware targeting end-users, governments, and business organizations. It has become a very profitable business for cybercriminals with revenues of millions of dollars, and a very serious threat to organizations with financial loss of billions of dollars. Numerous studies were proposed to address the ransomware threat, including surveys that cover certain aspects of ransomware research. However, no study exists in the literature that gives the complete picture on ransomware and ransomware defense research with respect to the diversity of targeted platforms. Since ransomware is already prevalent in PCs/workstations/desktops/laptops, is becoming more prevalent in mobile devices, and has already hit IoT/CPS recently, and will likely grow further in the IoT/CPS domain very soon, understanding ransomware and analyzing defense mechanisms with respect to target platforms is becoming more imperative. In order to fill this gap and motivate further research, in this paper, we present a comprehensive survey on ransomware and ransomware defense research with respect to PCs/workstations, mobile devices, and IoT/CPS platforms. Specifically, covering 137 studies over the period of 1990-2020, we give a detailed overview of ransomware evolution, comprehensively analyze the key building blocks of ransomware, present a taxonomy of notable ransomware families, and provide an extensive overview of ransomware defense research (i.e., analysis, detection, and recovery) with respect to platforms of PCs/workstations, mobile devices, and IoT/CPS. Moreover, we derive an extensive list of open issues for future ransomware research. We believe this survey will motivate further research by giving a complete picture on state-of-the-art ransomware research.


\end{abstract}

\ccsdesc[500]{Security and privacy~Malware and its mitigation}

\maketitle

\section{Introduction}

Recent years have witnessed a dramatic growth in the number of incidents a unique malware strain involved in, namely \emph{ransomware}. This notorious malware strain has been targeting not only ordinary end-users, but also governments and business organizations in almost any sector. Numerous incidents include Fortune 500 companies~\cite{Incident:Fortune500}, 
banks~\cite{Incident:BancoEstado}, cloud providers~\cite{Incident:BlackBoud}, chip manufacturers~\cite{Incident:TowerSemi}, cruise operators~\cite{Incident:Carnival}, threat monitoring services~\cite{Incident:ConnectWise}, governments~\cite{atlanta,Incident:Argentinia}, medical centers and hospitals~\cite{Incident:UHS}, schools~\cite{Incident:CarolinaSchool}, universities~\cite{Incident:Northumbria}, and even police departments~\cite{Incident:SwanseaPolice}. It has been predicted that the total loss of organizations due to ransomware will be around \$20 billion in 2021, and a new organization will be hit by those attacks in every 11 seconds~\cite{IncidentCosts}. Worse than that, in 2020, the first loss of human life as a result of ransomware attacks was reported to take place in Germany~\cite{Ransomware:Death}. Aforementioned incidents have already made ransomware the number one arms race problem between the threat and defense actors (i.e., governments, industry, and academia).



Ransomware (\emph{ransom} soft\emph{ware}) is a subset of malware designed to restrict access to a system or data until a requested ransom amount from the attacker is satisfied~\cite{symantecEvolution}. Based on the employed methodology, ransomware is generally classified into two types, namely \emph{cryptographic ransomware} that encrypts the victim files, and \emph{locker ransomware} that prevents victims from accessing to their systems. Regardless of the used methodology, both variants of ransomware demand a ransom payment to release the files or access to the system.
Although the first ransomware emerged in 1989 and has been intermittently around over 30 years, it has been one of the most notorious threats since 2005~\cite{cuttingGordian}.
Cybercriminals have perfected ransomware attack components (e.g., stronger encryption techniques, pseudo-anonymous payment methods, worm-like capabilities, etc.), and even started to serve ransomware as a service (RaaS)~\cite{Popoola:2017} by learning from past experiences and utilizing technological advancements over the time. 

A myriad of analysis, detection, and defense studies exist in the literature to address the ransomware threat. Several surveys were proposed that focus on certain aspects of ransomware research. However, no study exist in literature that covers evolution, characteristics, attack phases of ransomware as well as the complete picture on ransomware defense research by focusing on multitude of platforms (i.e., PCs/workstations, mobile devices, and IoT/CPS platforms). We believe that this is an important research gap in the literature since understanding the key characteristics of ransomware and existing defense solutions is becoming more and more crucial in combating this ever-growing threat. Since ransomware is already prevalent in PCs/workstations/desktops/laptops, is becoming more prevalent in mobile devices, and has already hit IoT/CPS recently, and will likely grow further in the IoT/CPS domain very soon, understanding ransomware and analyzing defense mechanisms with respect to target platforms is becoming more imperative. In order to fill this research gap, we present a comprehensive survey on ransomware and ransomware defense solutions with respect to PCs/workstations, mobile devices, and IoT/CPS platforms. Our survey covers 137 studies published in various conferences or journals over the period of 1989-2020. To the best of our knowledge, this is the first study in the literature that comprehensively analyzes the evolution of ransomware, draws a taxonomy of ransomware, and surveys the state-of-the-art ransomware defense research (i.e., analysis, detection, and recovery) with respect to various platforms (i.e., PCs/workstations, mobile devices, IoT/CPS environments).

\noindent\textbf{Contributions:} Contributions of this survey are listed as follows:
\begin{itemize}
    \item A detailed overview of ransomware evolution starting from 1989 to 2020 with respect to building blocks of ransomware and emergence of notable ransomware families.
    \item A comprehensive analysis of ransomware, key building blocks and their characteristics, and a taxonomy of notable ransomware families.
    \item An extensive overview of ransomware defense research (i.e., ransomware analysis, ransomware detection, and ransomware recovery) with a focus on multitude of platforms (PCs/workstations, mobile devices and IoT/CPS platforms).
    \item Derivation of a voluminous list of open research problems that need to be addressed in future ransomware defense research and practice. 
\end{itemize}

\noindent\textbf{Organization:} The structure of this survey is organized as follows: Section~\ref{sec:relatedwork} gives the related work. Section~\ref{sec:ransomware-and-evolution} provides an overview of ransomware and its evolution. Section~\ref{sec:taxonomy-ransomware} analyzes the key building blocks of ransomware and presents a taxonomy of notable ransomware families. Section~\ref{sec:taxonomy-research} gives an extensive overview of ransomware defense research with respect to PCs/workstations, mobile devices and IoT/CPS platforms. Section~\ref{sec:lessons} presents the open research problems that need to be addressed in future ransomware defense research. Section~\ref{section:conclusion} concludes the paper.


\section{RELATED WORK}\label{sec:relatedwork}
\begin{table*}[t!]
\caption{Comparison of the related work.}
\vspace{-1em}
\begin{tablenotes}
       \footnotesize
      \item \tikz\path pic{circle mark= 0}; = No information provided, \tikz\path pic{circle mark= 50}; = Partial information provided, \tikz\path pic{circle mark= 100}; = Complete information provided
    \end{tablenotes}
\label{table:comparison}
\begin{threeparttable}
\centering
\resizebox{1\textwidth}{!}{
\begin{tabular}{|c|p{8cm}|c|c|c|c|c|c|c|c|}
\hline
\multirow{2}{*}{\textbf{Work}} & \multirow{2}{*}{\textbf{\harun{Description}}} & 
\multirow{2}{*}{\textbf{Evolution}}
& \multicolumn{4}{c|}{\textbf{\harun{Covered} Characteristics of Ransomware}} 
& \multicolumn{3}{c|}{\textbf{Covered Platforms}}  \tabularnewline
\cline{4-10} & & &\textbf{Targets}& \textbf{Infection}&\begin{tabular}[c]{@{}l@{}}\textbf{Malicious} \\ \textbf{Actions}\end{tabular}& \textbf{Extortion} & \begin{tabular}[c]{@{}l@{}}\textbf{PCs/} \\ \textbf{Workstations}\end{tabular} & \begin{tabular}[c]{@{}l@{}}\textbf{Mobile} \\ \textbf{Devices}\end{tabular} & \begin{tabular}[c]{@{}l@{}}\textbf{IoT/} \\ \textbf{CPS}\end{tabular} \\ \hline
 
 Alzahrai et al. \cite{alzahrani:2017}   &\harun{Overview of ransomware in the Windows platform}&\tikz\path pic{circle mark= 0};       & \tikz\path pic{circle mark= 0}; & \tikz\path pic{circle mark= 0};    & \tikz\path pic{circle mark= 50};   & \tikz\path pic{circle mark= 0};   & \tikz\path pic{circle mark= 50};          & \tikz\path pic{circle mark= 0}; & \tikz\path pic{circle mark= 0};  \\ \hline            
 Aurangzeb et al. \cite{Aurangzeb:2017} &\harun{Survey on ransomware and trends}&\tikz\path pic{circle mark= 0};        & \tikz\path pic{circle mark= 50}; & \tikz\path pic{circle mark= 100};    & \tikz\path pic{circle mark= 100};   & \tikz\path pic{circle mark= 1000};   & \tikz\path pic{circle mark= 0};      & \tikz\path pic{circle mark= 0}; & \tikz\path pic{circle mark= 0}; \\ \hline            
 Mohan et al. \cite{KumarAndroid:2018} &\harun{Survey on the efficacy of Android ransomware detection techniques}&\tikz\path pic{circle mark= 0};        & \tikz\path pic{circle mark= 0}; & \tikz\path pic{circle mark= 0};    & \tikz\path pic{circle mark= 0};   & \tikz\path pic{circle mark= 0};   & \tikz\path pic{circle mark= 0};      & \tikz\path pic{circle mark= 100}; & \tikz\path pic{circle mark= 0}; \\ \hline     
 Abraham et al. \cite{abraham:2019}  &\harun{Survey on ransomware prevention using machine learning}&\tikz\path pic{circle mark= 0};        & \tikz\path pic{circle mark= 0}; & \tikz\path pic{circle mark= 0};    & \tikz\path pic{circle mark= 0};   & \tikz\path pic{circle mark= 0};   & \tikz\path pic{circle mark= 50};      & \tikz\path pic{circle mark= 0}; & \tikz\path pic{circle mark= 0}; \\ \hline     
 Dargahi et al. \cite{dargahi:2019}  &\harun{Cyber-Kill-Chain-based taxonomy of cryptographic ransomware}&\tikz\path pic{circle mark= 0};        & \tikz\path pic{circle mark= 0}; & \tikz\path pic{circle mark= 100};    & \tikz\path pic{circle mark= 100};   & \tikz\path pic{circle mark= 100};   & \tikz\path pic{circle mark= 100};      & \tikz\path pic{circle mark= 50}; & \tikz\path pic{circle mark= 50}; \\ \hline     
Maigada et al. \cite{maigida:2019}    &\harun{Review and metadata analysis of ransomware and defenses}&\tikz\path pic{circle mark= 0};        & \tikz\path pic{circle mark= 0}; & \tikz\path pic{circle mark= 0};    & \tikz\path pic{circle mark= 0};   & \tikz\path pic{circle mark= 0};   & \tikz\path pic{circle mark= 100};      & \tikz\path pic{circle mark= 100}; & \tikz\path pic{circle mark= 0}; \\ \hline     
 Keshavarzi et al. \cite{Keshavarzi:2020}  &\harun{Attack chain for ransomware offenses}& \tikz\path pic{circle mark= 0};        & \tikz\path pic{circle mark= 0}; & \tikz\path pic{circle mark= 100};    & \tikz\path pic{circle mark= 100};   & \tikz\path pic{circle mark= 100};   & \tikz\path pic{circle mark= 100};      & \tikz\path pic{circle mark= 100}; & \tikz\path pic{circle mark= 50}; \\ \hline  
Kok et al.~\cite{Kok2019RansomwareT} &\harun{Review of ransomware and detection techniques}&\tikz\path pic{circle mark= 0};        & \tikz\path pic{circle mark= 0}; & \tikz\path pic{circle mark= 50};    & \tikz\path pic{circle mark= 50};   & \tikz\path pic{circle mark= 0};   & \tikz\path pic{circle mark= 50};      & \tikz\path pic{circle mark= 0}; & \tikz\path pic{circle mark= 0}; \\ \hline  
 David et al. \cite{alzahraniDeep:2019}  &\harun{Review of Android ransomware detection using deep learning}& \tikz\path pic{circle mark= 0};        & \tikz\path pic{circle mark= 0}; & \tikz\path pic{circle mark= 0};    & \tikz\path pic{circle mark= 0};   & \tikz\path pic{circle mark= 0};   & \tikz\path pic{circle mark= 50};      & \tikz\path pic{circle mark= 0}; & \tikz\path pic{circle mark= 0}; \\\hline  
 Popoola et al.  \cite{Popoola:2017} &\harun{Ransomware trends, challenges, research directions}& \tikz\path pic{circle mark= 0};        & \tikz\path pic{circle mark= 0}; & \tikz\path pic{circle mark= 50};    & \tikz\path pic{circle mark= 50};   & \tikz\path pic{circle mark= 0};   & \tikz\path pic{circle mark= 50};      & \tikz\path pic{circle mark= 0}; & \tikz\path pic{circle mark= 50}; \\ \hline 
 Berrueta et al. \cite{berrueta:2020}  &\harun{Survey on cryptographic ransomware detection techniques}& \tikz\path pic{circle mark= 50};        & \tikz\path pic{circle mark= 50}; & \tikz\path pic{circle mark= 50};    & \tikz\path pic{circle mark= 100};   & \tikz\path pic{circle mark= 0};   & \tikz\path pic{circle mark= 100};      & \tikz\path pic{circle mark= 0}; & \tikz\path pic{circle mark= 0}; \\ \hline  
 Silva et al. \cite{silva:2019}  &\harun{Survey on situational awareness of ransomware attacks, detection, and prevention}& \tikz\path pic{circle mark= 0};        & \tikz\path pic{circle mark= 0}; & \tikz\path pic{circle mark= 50};    & \tikz\path pic{circle mark= 50};   & \tikz\path pic{circle mark= 0};   & \tikz\path pic{circle mark= 100};      & \tikz\path pic{circle mark= 100}; & \tikz\path pic{circle mark= 0}; \\ \hline  
 Bijitha et al.  \cite{bijitha:2020}&\harun{Survey on ransomware detection techniques}& \tikz\path pic{circle mark= 0};        & \tikz\path pic{circle mark= 0}; & \tikz\path pic{circle mark= 0};    & \tikz\path pic{circle mark= 0};   & \tikz\path pic{circle mark= 0};   & \tikz\path pic{circle mark= 50};      & \tikz\path pic{circle mark= 50}; & \tikz\path pic{circle mark= 0}; \\ \hline  
 Bajpai et al. \cite{keymanagement} &\harun{Key management-based taxonomy of ransomware}& \tikz\path pic{circle mark= 0};        & \tikz\path pic{circle mark= 0}; & \tikz\path pic{circle mark= 0};    & \tikz\path pic{circle mark= 100};   & \tikz\path pic{circle mark= 0};   & \tikz\path pic{circle mark= 0};      & \tikz\path pic{circle mark= 0}; & \tikz\path pic{circle mark= 0}; \\  \hline  
 Shinde et al. \cite{Shinde:2016}  &\harun{Study on ransomware transfer and mitigation}& \tikz\path pic{circle mark= 0};        & \tikz\path pic{circle mark= 0}; & \tikz\path pic{circle mark= 0};    & \tikz\path pic{circle mark= 50};   & \tikz\path pic{circle mark= 0};   & \tikz\path pic{circle mark= 0};      & \tikz\path pic{circle mark= 0}; & \tikz\path pic{circle mark= 0}; \\ \hline  
 Gonzalez et al. \cite{Gonzales:2017} &\harun{Detection and prevention of cryptographic ransomware}& \tikz\path pic{circle mark= 0};        & \tikz\path pic{circle mark= 0}; & \tikz\path pic{circle mark= 100};    & \tikz\path pic{circle mark= 100};   & \tikz\path pic{circle mark= 0};   & \tikz\path pic{circle mark= 50};      & \tikz\path pic{circle mark= 0}; & \tikz\path pic{circle mark= 0}; \\ \hline  
Humayun et al. \cite{IoTRans} &\harun{Ransomware evolution, mitigation, and prevention in IoT}& \tikz\path pic{circle mark= 0};        & \tikz\path pic{circle mark= 0}; & \tikz\path pic{circle mark= 100};    & \tikz\path pic{circle mark= 0};   & \tikz\path pic{circle mark= 0};   & \tikz\path pic{circle mark= 0};      & \tikz\path pic{circle mark= 0}; & \tikz\path pic{circle mark= 100}; \\ \hline  
 Al-rimy et al. \cite{threatfactors} &\harun{Survey on ransomware success factors, taxonomy, and defenses}& \tikz\path pic{circle mark= 0};        & \tikz\path pic{circle mark=50}; & \tikz\path pic{circle mark= 100};    & \tikz\path pic{circle mark= 100};   & \tikz\path pic{circle mark= 0};   & \tikz\path pic{circle mark= 100};      & \tikz\path pic{circle mark= 0}; & \tikz\path pic{circle mark= 0}; \\ \hline  
Garg et al. \cite{Garg2018APE} &\harun{Past and future of ransomware}& \tikz\path pic{circle mark= 50};        & \tikz\path pic{circle mark= 0}; & \tikz\path pic{circle mark= 100};    & \tikz\path pic{circle mark= 0};   & \tikz\path pic{circle mark= 0};   & \tikz\path pic{circle mark= 50};      & \tikz\path pic{circle mark= 0}; & \tikz\path pic{circle mark= 0}; \\ \hline  
Alzahrani et al. \cite{Alzahrani:2020}  &\harun{Ransomware in Windows and Android platforms}& \tikz\path pic{circle mark= 50};        & \tikz\path pic{circle mark= 0}; & \tikz\path pic{circle mark= 50};    & \tikz\path pic{circle mark= 50};   & \tikz\path pic{circle mark= 50};   & \tikz\path pic{circle mark= 50};      & \tikz\path pic{circle mark= 50}; & \tikz\path pic{circle mark= 0}; \\ \hline  

     
Naseer et al. \cite{Naseer:2020}&\harun{Survey on Windows ransomware}& \tikz\path pic{circle mark= 0};        & \tikz\path pic{circle mark= 50}; & \tikz\path pic{circle mark= 50};    & \tikz\path pic{circle mark= 0};   & \tikz\path pic{circle mark= 0};   & \tikz\path pic{circle mark= 50};      & \tikz\path pic{circle mark= 0}; & \tikz\path pic{circle mark= 0}; \\ \hline  
     
Zimba et al. \cite{zimba:2019}  &\harun{Evolution of ransomware}& \tikz\path pic{circle mark= 100};        & \tikz\path pic{circle mark= 50}; & \tikz\path pic{circle mark= 0};    & \tikz\path pic{circle mark= 100};   & \tikz\path pic{circle mark= 0};   & \tikz\path pic{circle mark= 0};      & \tikz\path pic{circle mark= 0}; & \tikz\path pic{circle mark= 0}; \\ \hline  
Rehman et al. \cite{Rehman:2018}  &\harun{Security assurance against ransomware}& \tikz\path pic{circle mark= 50};        & \tikz\path pic{circle mark= 50}; & \tikz\path pic{circle mark= 50};    & \tikz\path pic{circle mark= 50};   & \tikz\path pic{circle mark= 0};   & \tikz\path pic{circle mark= 0};      & \tikz\path pic{circle mark= 0}; & \tikz\path pic{circle mark= 0}; \\ \hline  

Ibarra et al. \cite{scadaCPS}&\harun{Impact of ransomware on SCADA systems}& \tikz\path pic{circle mark= 0};        & \tikz\path pic{circle mark= 0}; & \tikz\path pic{circle mark= 50};    & \tikz\path pic{circle mark= 0};   & \tikz\path pic{circle mark= 0};   & \tikz\path pic{circle mark= 0};      & \tikz\path pic{circle mark= 0}; & \tikz\path pic{circle mark= 50}; \\ \hline  
          
Kiru et al. \cite{Kiruetal.}  &\harun{Understanding ransomware and countermeasures}& \tikz\path pic{circle mark= 0};        & \tikz\path pic{circle mark= 50}; & \tikz\path pic{circle mark= 50};    & \tikz\path pic{circle mark= 50};   & \tikz\path pic{circle mark= 50};   & \tikz\path pic{circle mark= 50};      & \tikz\path pic{circle mark= 0}; & \tikz\path pic{circle mark= 0}; \\ \hline  
Desai et al.  \cite{Desai_2018}  &\harun{Survey on Android ransomware and detection methods}& \tikz\path pic{circle mark= 50};        & \tikz\path pic{circle mark= 50}; & \tikz\path pic{circle mark= 50};    & \tikz\path pic{circle mark= 0};   & \tikz\path pic{circle mark= 50};   & \tikz\path pic{circle mark= 50};      & \tikz\path pic{circle mark= 0}; & \tikz\path pic{circle mark= 0}; \\\hline  
     
Lipovsky et al. \cite{TheriseofAndroid}  &\harun{The rise of Android ransomware}& \tikz\path pic{circle mark= 50};        & \tikz\path pic{circle mark= 50}; & \tikz\path pic{circle mark= 50};    & \tikz\path pic{circle mark= 0};   & \tikz\path pic{circle mark= 50};   & \tikz\path pic{circle mark= 50};      & \tikz\path pic{circle mark= 0}; & \tikz\path pic{circle mark= 0}; \\ \hline  
     
Maniath et al. \cite{mainath:2018}&\harun{Survey on prevention, mitigation, and containment of ransomware}& \tikz\path pic{circle mark= 50};        & \tikz\path pic{circle mark= 50}; & \tikz\path pic{circle mark= 50};    & \tikz\path pic{circle mark= 0};   & \tikz\path pic{circle mark= 50};   & \tikz\path pic{circle mark= 50};      & \tikz\path pic{circle mark= 0}; & \tikz\path pic{circle mark= 0}; \\ \hline  
\rowcolor[rgb]{ .863,  .902,  .945}\textbf{This work}&\harun{\textbf{Comprehensive survey on ransomware evolution, taxonomy, and defenses in PCs/workstations, mobile devices, and IoT/CPS}}& \tikz\path pic{circle mark= 100};        & \tikz\path pic{circle mark= 100}; & \tikz\path pic{circle mark= 100};    & \tikz\path pic{circle mark= 100};   & \tikz\path pic{circle mark= 100};   & \tikz\path pic{circle mark= 100};      & \tikz\path pic{circle mark= 100}; & \tikz\path pic{circle mark= 100}; \\ \hline

\end{tabular}}
\end{threeparttable}
\end{table*}

Ransomware has been a very active topic of research, and several researchers proposed surveys that focus on different aspects of ransomware research. 

\vspace{0.2em}

\noindent\textbf{Ransomware for PCs/workstations.} Aurangzeb et al.~\cite{Aurangzeb:2017} summarized the current trends of ransomware for PCs. A short overview of ransomware and mitigation strategies was given in~\cite{Gonzales:2017}. 
Popoola et al.~\cite{Popoola:2017} provided an overview of both successful and unsuccessful ransomware strains. Garg et al.~\cite{Garg2018APE} discussed the infection methods, prevention measures, and future of ransomware. 
Rehman et al.~\cite{Rehman:2018} gave a short overview WannaCry ransomware. Shinde et al.~\cite{Shinde:2016} and Kiru et al.~\cite{Kiruetal.} discussed the underlying success of ransomware attacks. 
Maigida et al.~\cite{maigida:2019} provided a review of metadata analysis of ransomware attacks. Bajpai et al.~\cite{keymanagement} provided a taxonomy of ransomware based on key management techniques. Zimba and Chishimba~\cite{zimba:2019} categorized ransomware strains based on encryption and deletion processes. The works~\cite{dargahi:2019,Keshavarzi:2020} 
analyzed attack phases of ransomware based on Cyber-Kill-Chain, and attack channel models respectively. Considering the ransomware defense solutions for PCs, 
Alzahrani et al. \cite{alzahrani:2017} focused on the ransomware 
defenses for the Windows platform. The works presented in~\cite{abraham:2019,alzahraniDeep:2019} gave an overview of the defenses that use Machine Learning (ML) and Deep Learning (DL). 
The works \cite{bijitha:2020, berrueta:2020,silva:2019,Kok2019RansomwareT,mainath:2018,threatfactors} surveyed the ransomware   
defense solutions. 

\vspace{0.2em}

\noindent\textbf{Ransomware for Mobile Devices.} 
The works proposed in~\cite{TheriseofAndroid,Desai_2018,KumarAndroid:2018} reviewed the ransomware research for mobile platforms. Lipovský et al.~\cite{TheriseofAndroid} analyzed the evolution and behavior of Android ransomware. Desai summarized the ransomware analysis techniques for Android platforms~\cite{Desai_2018}. Lastly, Kumar et al.~\cite{KumarAndroid:2018} reviewed the ransomware detection techniques for Android platforms. In terms of the studies focusing on both PCs and mobile devices, Alzahrani et al.~\cite{Alzahrani:2020} surveyed the evolution, strains, analysis and defense techniques in both Windows and Android platforms.

\vspace{0.2em}

\noindent\textbf{Ransomware for IoT/CPS Platforms.} 
Only a few works exist in the literature that focus on the IoT/CPS ransomware. 
Humayun et al.~\cite{IoTRans} examined the evolution of ransomware on IoT platforms. Ibarra et al.~\cite{scadaCPS} discussed the efficacy of ransomware on the CPS environments, and categorized the ransomware defense solutions.

\vspace{0.2em}

\noindent\textbf{Differences from existing surveys:} The main differences of our work from the prior works are as follows: (1) Existing works did not give a comprehensive view of the evolution of ransomware. In contrast, we comprehensively analyze the evolution of ransomware and notable events in the ransomware evolution as it is crucial to understand historical technical trends in ransomware. (2) Most of the surveys focused only on the specific phases of ransomware attacks (e.g., infection). On the other hand, we extensively analyze every attack phases of ransomware. (3) While prior works briefly summarized the defense solutions for a single platform, we analyze the defense solutions (i.e., analysis, detection, and recovery) for the majority of platforms such as PCs/workstations, mobile devices, and IoT/CPS environments. The comparison of our survey against the existing surveys is outlined in Table~\ref{table:comparison}. To the best of our knowledge, this is the most comprehensive survey in the literature as of the time of writing this paper.

\section{RANSOMWARE AND EVOLUTION OF RANSOMWARE}\label{sec:ransomware-and-evolution}

Ransomware is a subset of malware that prevents or limits users from accessing their system and/or data until a ransom is paid~\cite{kirdaBlackHat}. The main objective of ransomware is extorting money from the victims. Based on the employed methodology, ransomware is generally classified into two types. 

\vspace{0.4em}
\noindent\emph{Cryptographic Ransomware:} This variety of ransomware 
encrypts victim files, deletes or overwrites the original files, and demands a ransom payment for decryption of the files. 

\vspace{0.4em}
\noindent\emph{Locker Ransomware:} This type of ransomware prevents the victim from accessing its system by locking the screen or browser, and demands a ransom payment to unlock the system. Unlike cryptographic ransomware, it does not encrypt the system or user data.

\begin{figure}[htbp]
\centering
\includegraphics[scale=0.55]{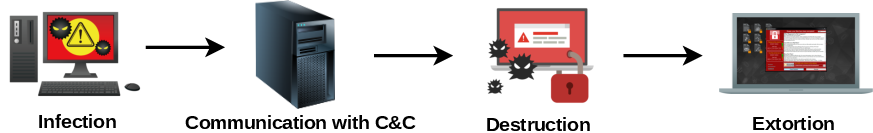}   
\caption{\harun{Generalized overview of attack phases of ransomware in which items in the model build upon \cite{alzahrani:2017,threatfactors,berrueta:2020,Keshavarzi:2020,maigida:2019}.}}
\label{fig:phases}
\vspace{-1em}
\end{figure}

A generalized overview of ransomware attack phases is shown in Figure~\ref{fig:phases} \harun{which we build upon prior studies \cite{alzahrani:2017,threatfactors,berrueta:2020,Keshavarzi:2020,maigida:2019}. Although some ransomware may not possess an individual phase in the shown model, such as Communication with C\&C, our model here in this work generalizes the attack phases of ransomware}. Attack phases of ransomware can be summarized as follows: 
\begin{itemize}
    \item \emph{Infection:} In this phase, ransomware is delivered to a victim system (e.g., PC/workstation, mobile device, IoT/CPS device, etc.). Malicious actors employ several infection vectors to achieve the delivery of ransomware.
    \item \emph{Communication with C\&C servers:} After the infection, ransomware connects to the Command and Control (C\&C) server to exchange crucial information (i.e., encryption keys, target system information) with the attacker. Although several ransomware strains communicate with C\&C servers, there exist some families that do not perform any communication.
    \item \emph{Destruction:} In this phase, ransomware performs the actual malicious actions such as encrypting files or locking systems to prevent the access of the victim to his/her files or system. 
    \item \emph{Extortion:} Finally, the ransomware informs the victim about the attack by displaying a ransom note. The ransom note discloses the attack details and payment instructions. 
\end{itemize}
 
We note that some ransomware families display worm-like behavior, in which they try to infect more victims that reside in the same network. We analyze each attack phase in further detail in Section~\ref{sec:taxonomy-ransomware}. However, before that, we comprehensively dig into the evolution of ransomware where we point out important events from the emergence of ransomware until 2020.

\subsection{Evolution of Ransomware}\label{subsec:evolution}
\begin{figure}[!t]
\centering
\includegraphics[width=\linewidth]{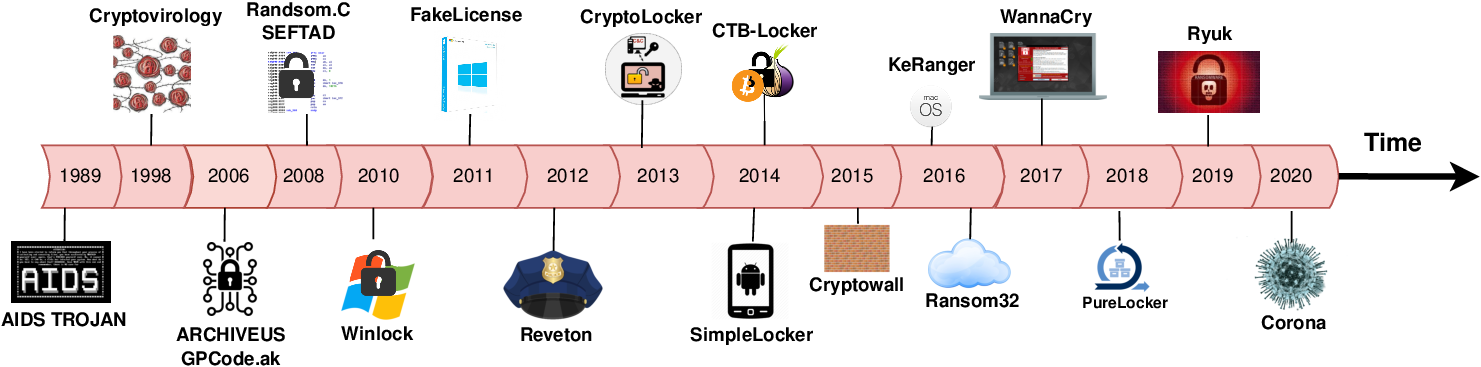}
\vspace{-0.75em}
\caption{Evolution of Major Ransomware Families from 1989 to 2020.}
\label{fig:evolution}
\vspace{-2em}
\end{figure}

Although ransomware attacks immensely increased in the last decade, the history of ransomware almost begins with the emergence of the first PCs. 
The evolution of ransomware considering the milestones is shown in Figure~\ref{fig:evolution}.

\emph{The first ransomware} 
- \emph{AIDS Trojan (aka, PC Cyborg)} was created in 1989~\cite{bates1990trojan}. 20,000 infected floppy disks were distributed to the attendees of the 
AIDS conference by mail. It was encrypting file names on the C:\textbackslash{ } drive of the infected computer with a custom symmetric encryption algorithm, and demanding a ransom. 
Seven years after this incident, 
researchers explained the faults of the PC Cyborg and outlined the emergence of a new \emph{cryptovirology} concept~\cite{young96}.  
They developed a proof-of-concept (PoC) malware that uses public key cryptography to encrypt the user data~\cite{LYoung:2017} to caution the community about the future digital extortion crimes. 


Apart from the AIDS Trojan and the cryptovirology, ransomware remained 
silent 
until 2005 probably due to 
the yet underdeveloped information technology infrastructure, scarcity of the Internet connectivity, and infrequency of the world-wide-web (WWW). 
However, the Internet and WWW got more prevalent; social media, blogging and e-commerce platforms emerged, and the number of users connected to the Internet reached one billion by 2005 
which brought back the digital extortion~\cite{Argaez}, and \emph{GPCode} - the first modern cryptographic ransomware emerged. \emph{GPCode} was infecting the target computers via phishing emails, 
using a custom symmetric encryption algorithm, and storing the encryption key at the victim side. Although 
it was ineffective, 
it provided an example design pattern 
for 
future ransomware~\cite{Nikolai_2016}. 

Between 2005 and 2006, 
\emph{CryZip}, \emph{Archiveus}~\cite{Archiveus}, and \emph{Krotten}~\cite{Krotten} emerged as the earliest ransomware families that utilized asymmetric encryption. 
Usage of public and private keys for encryption and decryption processes was a momentous step for ransomware, 
and made the recovery attempts almost impossible without knowing the attacker's decryption key. 


\emph{The first locker ransomware - Randsom.C} appeared in 2008~\cite{symantecEvolution}. 
It was locking the victim's desktop and displaying a ransom message that claims to be from Windows Security Center, and asking the user to call a premium-rate phone number to reactivate the license~\cite{symantecEvolution}. In the same year, \emph{Seftad} ransomware heralded with a new method of modifying target computer's Master Boot Record (MBR) to prevent the system from booting normally~\cite{seftad1}. Then, it was asking for a ransom 
via prepaid payment method such as Paysafecard~\cite{paysafecard}. 

Up until the \emph{emergence of cryptocurrencies}, the 
major bottleneck for ransomware was 
the ransom payment. 
There was no approach for ransomware authors that does not limit the payments to certain geographies, is not liable to local law authorities, and protects their anonymity yet allows the transfer of big amounts of ransoms~\cite{Huang:2018}. The emergence and prevalence of cryptocurrencies after 2009, such as Bitcoin, helped cybercriminals to solve these problems. Since attackers believed that their anonymity were preserved via blockchain (in fact blockchain transactions can be traced that make it pseudo-anonymous~\cite{Huang:2018,ahmetbitcoin1,ahmetbitcoin2,ahmetbitcoin3,ahmetBitcoin4}), ransomware was able to overcome the biggest operational bottleneck. This advancement led threat actors to carry out more widespread 
ransomware attacks. About 60,000 new ransomware families were detected in 2011~\cite{ransomwareAttacks}.

Another notable locker ransomware \emph{Reveton} (aka \emph{Police ransomware}) showed up with a different technique in 2012. 
In addition to locking the victim's computer, 
it was trying to exfiltrate valuable information from the victim's computer~\cite{Cimpanu_2019}. 
In the meantime, \emph{CryptoLocker} was born as an initiator of advanced cryptographic ransomware variants in 2013. It was 
encrypting certain file types (i.e., .pdf, .zip) using 2048-bit RSA and demanding ransom in Bitcoin. 

In 2014, \emph{Curve-Tor-Bitcoin (CTB) Locker} arrived which 
took its name based on the key technologies it was using. 
\emph{Curve} was signifying the use of Elliptic Curve Cryptography (ECC) for encryption, 
\emph{TOR} was representing the anonymity-preserving web browsing scheme to be used during ransom payment, and 
\emph{Bitcoin} was referring to the 
ransom payment~\cite{CTBLocker}. In the same year, 
\emph{Cryptowall} cryptographic ransomware emerged which 
was also using TOR and Bitcoin, and deleting volume shadow copies to prevent the restoration of the files. 
It infected more than 600,000 systems~\cite{Cryptowall}. 

The first mobile locker ransomware, namely \emph{Android Defender} arrived in 2014. It was 
tricking users by disguising itself as a legitimate antivirus application~\cite{TheriseofAndroid}. One year later, the first mobile cryptographic ransomware - Android Defender emerged.  
After infection, it was scanning mobile device's SD card and encrypting files with specific extensions using AES. The hard-coded encryption key  in the binary 
made it trivial to extract the key to 
decrypt the files~\cite{mobileRansomware}.



Starting from 2015, ransomware began to target 
other 
operating systems. In 2015 \emph{Linux.Encoder}~\cite{Bisson_2015} appeared as the first ransomware targeting GNU/Linux platforms~\cite{DoctorWeb_2015}. 
It was encrypting the \emph{home} directory and directories related to website administration. 
Next year, the first macOS ransomware \emph{KeRanger} 
was signed with a valid Mac app development certificate to bypass Apple's protection mechanism. 
Both Linux.Enconder and KeRanger were using the hybrid encryption~\cite{keranger2016}. 


As a new business model on cybercrime, the threat of ransomware moved to a new dimension by the emergence of \emph{Ransomware-as-a-Service (RaaS)} in 2015. RaaS aimed to provide user-friendly, and easy-to-modify ransomware kits that could be purchased by anyone on underground markets. 
That was a momentous step for the evolution of ransomware as it could be easily repackaged to infect any platform which made it platform-agnostic. RaaS escalated the number of ransomware attacks around the world~\cite{raas}.


In 2017, \emph{WannaCry} ransomware appeared and 
became the 
the worst cybercrime of that year. It affected more than 250.000 systems in 150 countries~\cite{wannacry} with the help of the Microsoft Windows SMB Server Remote Code Execution Vulnerability. It used 
AES to encrypt each file with a different key, then individual keys were encrypted using a 2048-bit RSA~\cite{wannacry}. 

In 2018 \emph{PureLocker} appeared that was written in PureBasic programming language making it platform-agnostic. It was using hybrid encryption 
and displaying a ransom note in which the attacker was requesting victims to contact him/her via Proton untraceable secure email service. In the recent years, cybercriminals started to design new ransomware families that target specific victims. One such example is \emph{Ryuk}, seen in 2019, which was 
targeting only enterprises~\cite{Incident:TowerSemi}. Unlike other ransomware, Ryuk was mostly infecting its targets via other malware, most notably TrickBot. 

During the global pandemic in 2020, the need for health centers, thus their vulnerabilities, 
increased the number of ransomware attacks to health organizations, and even a new ransomware strain named \emph{Corona} emerged~\cite{Infostealer_2020}. Corona was targeting the hospitals and it was encrypting health records of patients. After that, it was displaying a COVID-19-themed ransom message. 

\harun{As it can be seen from the evolution of ransomware, this notorious threat started as a weak threat in 1989 lack of strong and fast encryption techniques, diverse infection vectors, (pseudo)anonymous payment methods, and a wide variety of targets. However, as the technology evolved, ransomware authors learned from prior unsuccessful attempts and technological advancements, hence achieved to make ransomware the number one cyber threat. Such an evolution left its impacts not only on end-users, but also on organizations, enterprises, and critical infrastructures. While it was possible for security researchers to recover the files/system successfully after the first examples of (unsuccessful) ransomware attacks, currently, it is almost impossible to recover the files/system without the ransom payment or restoration of available backups. Successful ransomware attacks not only cause their targets to lose money and time, but also to lose reputation. As ransomware is evolving from platform-dependent to platform-independent, and from simple ransomware to a fully-fledged RaaS model, it is becoming more and more prevalent, threatening almost every computerized system/target.}

\section{TAXONOMY OF RANSOMWARE}\label{sec:taxonomy-ransomware}

\begin{figure*}[htbp]
\centering
\includegraphics[width=\linewidth]{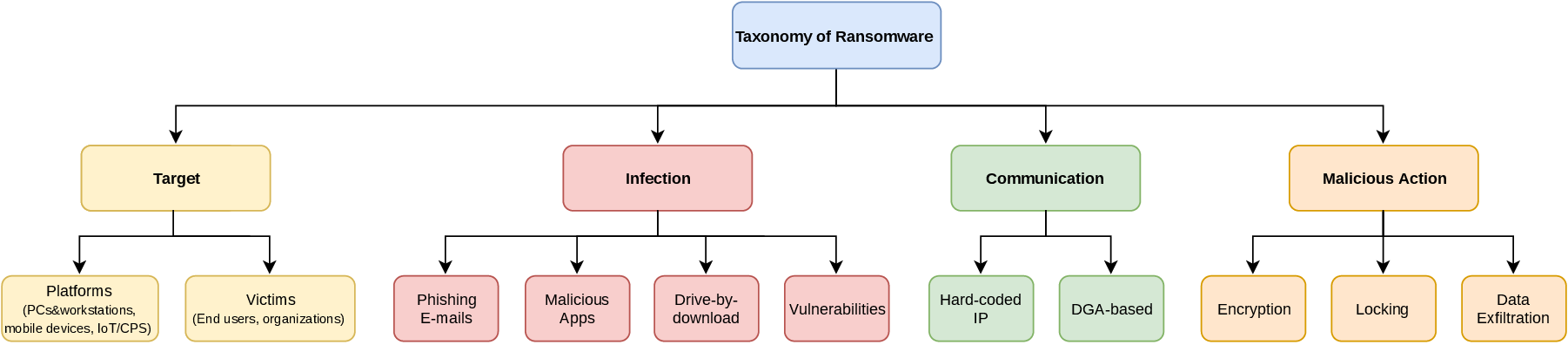}
\caption{Taxonomy of ransomware.}
\label{fig:methodology}
\end{figure*}

Ransomware can be classified in various ways. In this study, we classify ransomware with respect to their \emph{target, infection method, C\&C communication}, and \emph{malicious action (destruction technique)} as shown in Figure~\ref{fig:methodology}. In this section, we firstly provide an overview for each classification category, and then classify the notable ransomware families based on our methodology.

\subsection{Classification by Target}
Ransomware can be classified with respect to their targets under two categories that are orthogonal to each other: \emph{target victim} and \emph{target platform}. 

\subsubsection{Victims of Ransomware}\hfill 

Ransomware can target a variety of victim types. Analyzing the victim types of ransomware can provide valuable information towards designing practical defense mechanisms. Victims of ransomware can be divided into two groups: \emph{End-users} and \emph{Organizations}.

\emph{End-Users} were the primary targets for the first ransomware families. 
Lack of security awareness, and technical assistance make ransomware especially effective against end-users \cite{symantecEvolution}. Cryptographic ransomware can encrypt worth-to-pay files of individuals that are stored in the personal devices (e.g., PCs, laptops, smartphones, etc.). Meanwhile, locker variants may lock end-user's 
devices and prevent access unless a ransom amount is paid. Unsurprisingly, demanded ransom amount from end-users is significantly lower than the amount for organizational targets~\cite{symantecEvolution}. Moreover, a single ransomware may infect thousands of end-user systems, that makes 
it profitable~\cite{volentiVictim}. 

\emph{Organizations} were not initially the main targets of ransomware. However, as ransomware evolved in time, many types of organizations including 
governments, 
hospitals~\cite{ransomwareHospital}, enterprises, and schools~\cite{schoolRans} were targeted 
frequently. In those 
attacks, cybercriminals choose their targets in advance, and attempt to cause maximum disruption 
in the hope of a big ransom payment~\cite{Obrien_2017}. Locker ransomware can lock computers used in the target that may cause the organization's entire operation to stop~\cite{atlanta}. Likewise, cryptographic ransomware can encrypt valuable information stored in the organization's system, and make it inaccessible until a huge ransom amount is paid. Cybercriminals can also threaten to publish their target's data to the public.

\subsubsection{Target Platforms of Ransomware}\hfill 

Another significant point to understand the behavior of ransomware is the target platform. Ransomware targets a variety of platforms. Most of the time, it is specifically designed for a platform and an objective operating system because it often leverages the system-specific libraries/functions (i.e., system calls) to perform its malicious actions \cite{symantecEvolution}. In this study, we will use \emph{platform} and \emph{operating system} terms interchangeably, and divide the target platforms of ransomware into three groups: \emph{PCs/workstations}, \emph{mobile devices}, and \emph{IoT/CPS devices}.

\emph{PCs/workstations.} The most common targets of ransomware are PCs/workstations. Due to the popularity among users~\cite{windowsMarket}, the majority of ransomware 
target PCs and workstations with Windows OS. 
In addition, there are some ransomware families that target other operating systems, such as KeRanger for macOS, and LinuxEncoder for GNU/Linux platforms. 
The victims can mitigate screen locker ransomware attacks by re-installing their OS. On the other hand, concerning cryptographic ransomware, 
it is almost impossible to decrypt and recover the files due to utilization of advanced cryptography techniques~\cite{TANG2020101997}. So, cryptographic ransomware families are the main threats for PCs/workstations.


\emph{Mobile Devices.} 
The increasing popularity of mobile devices 
in society makes mobile devices such as smartphones ideal targets for ransomware. In 
terms of mobile devices, ransomware 
target Android and iOS platforms since these two platforms share the biggest global mobile 
OS market. 
Apple has a hard-controlled ecosystem where applications are thoroughly vetted before being published to customers. Probably for this reason, iOS users have not been affected by ransomware. There have been only fake ransomware examples for iOS devices~\cite{fakeios}. Quite the contrary, due to the open ecosystem of the Android platform, ransomware is a severe threat for Android users. In fact, the first locker ransomware for mobile devices, namely Android Defender - emerged in 2013, targeted Android platforms, and in the following year, the first cryptographic ransomware, Simplocker, emerged \cite{TheriseofAndroid}. Even though for PCs/workstations cryptographic ransomware are more threatening than locker variants, it is the opposite way for mobile ransomware. The underlying reason is that, the effect of locker ransomware on PCs/workstation can be avoided most of the time by removing the hard-drive \cite{Snow_2016} whereas on mobile devices, the same process is not easy.

\emph{IoT/CPS Devices.} IoT and CPS devices are not the major targets of ransomware strains at the moment. \ahmet{However, such devices are becoming more and more ubiquitous in numerous deployment areas including but not limited to smart homes, smart health, smart buildings, smart transportation, smart cities, smart factories, etc.~\cite{iqtidar, mekdad2021survey,yacoub}}. In fact, Industrial IoT and CPS devices (e,g, PLCs, RTUs, RIOs, etc.) have already been driving the industrial control systems in smart grids, water and gas pipes, and nuclear and chemical plants. Although the existing ransomware~\cite{IoT1} for such devices are not prevalent right now, adversaries can target such environments much more in the future.

\vspace{-0.5em}

\subsection{Classification by Infection Vectors}
 Ransomware authors employ the infection techniques that are used for traditional malware to infect their targets. Infection methods of ransomware can be categorized into five groups: \emph{malicious e-mails, SMS or instant messages (IMs), malicious applications, drive-by-download}, and \emph{vulnerabilities}.

\emph{Malicious e-mails} are the most commonly used infection vectors for ransomware. Attackers send spam e-mails to victims that have attachments containing ransomware~\cite{ODonnell_2019}. Such spam campaigns can be distributed using botnets~\cite{Obrien_2017,kurt2020lnbot, kurt2021lnbot}. Ransomware may come with an attached malicious 
file, or the e-mail may contain a malicious link that will trigger the installation of ransomware once visited (drive-by download). 

\emph{SMS Messages or IMs} are used frequently for mobile ransomware. In such kind of infections, attackers send SMS messages or IMs to the victims that will cause them to browse a malicious website to download ransomware to their platforms~\cite{ODonnell_2019,TheriseofAndroid}.

\emph{Malicious Applications} are used by ransomware attackers who develop and deploy mobile applications that contain ransomware camouflaged as a benign application~\cite{ODonnell_2019,TheriseofAndroid}.

\emph{Drive-by download} happens when a user unknowingly visits an infected website or clicks a malicious advertisement (i.e., malvertisement) and then the malware is downloaded and installed without the user’s knowledge~\cite{infection1}.

\emph{Vulnerabilities} in the victim platform such as vulnerabilities in operating systems~\cite{wannacry},  browsers~\cite{browser1}, or software can be used by ransomware authors as infection vectors. Attackers can use helper applications, 
\emph{exploit kits}, to exploit the known or zero-day vulnerabilities in target systems. Attackers can redirect victims to those kits via malvertisement and malicious links.

 



\subsection{Classification by C\&C Communication}
A command-and-control (C\&C) server is a remote server in the attacker's domain~\cite{Commandandcontrol}. C\&C servers are frequently used by adversaries to communicate and configure the malware. In the context of ransomware, C\&C servers are mainly used by cryptographic ransomware families to send or receive the encryption key that is used to encrypt the files and/or applications of the victim. Ransomware families mostly use HTTP or HTTPS protocols for this aim~\cite{protocol}. Ransomware families can connect to the C\&C server either via \emph{hard-coded IP addresses or domains}, or \emph{dynamically fast-fluxed/generated/shifted  domain names using Domain Generation Algorithms (DGA)}. 

\emph{Hard-coded IPs/Domains:} Ransomware families can embed hard-coded IP addresses or domains to their binaries 
to setup a connection to the C\&C server. In 
this approach, IP address or the domain remains the same for every attack, and provides a reliable communication for attackers. However, those hard-coded values can be used by defense systems to create signatures for detection. 

\emph{Dynamic Domains:} Domain Generation Algorithms (DGA) are used by ransomware families 
in order to contact C\&C servers dynamically. Those algorithms 
provide a unique domain name to the server for each communication by fast-fluxing/generating/shifting the domain names. 
This form of communication serves to communicate more robustly for ransomware, and firewalls cannot easily detect it~\cite{dgaBased}. 
\vspace{-0.75em}

\subsection{Classification by Malicious Action}
Even though all ransomware families are designed to extort money from their victims, they can show different characteristics in terms of their malicious actions. The malicious actions that can be taken by ransomware can be divided into two groups: \emph{encrypting} and \emph{locking}.
\vspace{-0.5em}
\subsubsection{Encrypting}\hfill

Encryption is a malicious action implemented by cryptographic ransomware families that aim to 
prevent access to victim files unless a ransom is paid. Ransomware first prepares the keys, and then starts 
the encryption process. 
Previously, ransomware families were solely encrypting the files located in the specific part of the hard drive \cite{bates1990trojan}. Over time, ransomware authors started to target specific file types (i.e., .doc, .zip, .pdf) that may contain valuable information of victims. After the encryption process, ransomware can display various destruction behavior on the original victim files, such as deleting or overwriting. In this subsection, we firstly explain the encryption techniques used by ransomware, and then give brief overview of destruction behaviors.

\vspace{0.4em}
\emph{Encryption Techniques:} Ransomware can employ \emph{symmetric}, \emph{asymmetric},  or \emph{hybrid} encryption techniques. To perform the encryption operation, ransomware can utilize system APIs, or pre-implemented encryption algorithms located in the actual source code of the ransomware~\cite{miningAPICall}.

\vspace{0.2em}
\noindent\emph{Symmetric-Key Encryption: } Only one key is used to encrypt and decrypt files in symmetric-key encryption. Compared to asymmetric-key encryption, 
it requires a lower amount of 
resources for the encryption of a large number of files so ransomware can encrypt victim files faster 
~\cite{crypto1}. However, 
the attacker needs to ensure that the key is inaccessible to the victim after the encryption process~\cite{symantecEvolution}. 
Encryption key is either generated 
at the target system, or embedded into the ransomware binary
. After the encryption, ransomware sends the encryption key to the attacker through C\&C communication. Although ransomware families have been using different symmetric-key encryption algorithms, AES (Advanced Encryption Standard) is the most popular algorithm. 

\vspace{0.2em}
\noindent\emph{Asymmetric-Key Encryption: } In this method, ransomware utilizes a pair of keys, namely public and private keys, to encrypt and decrypt files. 
Although 
not efficient to 
encrypt large number of files
, asymmetric-key encryption solves the key protection problem since separate keys are required for encryption and decryption. 
Attackers can embed public key into the binary as in TeslaCrypt~\cite{deployment} that allows ransomware to start encryption without connecting to the C\&C. They can also generate the keys on victim systems as in CryptoLocker \cite{Cannell_2013}. In some ransomware families, such as WannaCry~\cite{wannacry}, the attacker's public key is delivered through C\&C communication. So connection to the C\&C server is required to start encryption. Moreover, some variants can generate unique public-private key pairs for every victim. This allows the attacker to decrypt files on one victim without revealing the private key that could also be used to decrypt files on other victims
~\cite{symantecEvolution}. RSA (Rivest–Shamir–Adleman) is the most frequently used asymmetric key algorithm.

\vspace{0.2em}
\noindent\emph{Hybrid Encryption:}  Advantages of both of the encryption techniques are combined by attackers in hybrid encryption. In this respect, ransomware first uses symmetric key encryption to encrypt victim's files quickly. After that, it encrypts the used symmetric key with the attacker's public key. Generally, attacker's public key is embedded in the ransomware  binary, so that those variants do not require connection to the C\&C server during the attack.

\vspace{0.4em}
\emph{Destruction Behaviors:}
Ransomware can display different behaviors for destructing the victim's original files after completing the encryption process. Some ransomware families encrypt the files in-place such that they \emph{overwrite} the original file with the encrypted versions. On the other hand, some families delete original files of the victim by modifying the Master File Table (MFT), and create a new file that contains the encrypted version of the original file~\cite{cuttingGordian}. To eliminate the chance of restoration of the files from the file system snapshots, some ransomware strains such as Locky, delete Windows Volume Shadow copies after the infection~\cite{Locky}.

\subsubsection{Locking}\hfill 

Locker ransomware families lock system components to prevent the access of victims. Based on the 
target, locking ransomware can be divided into three categories: \emph{screen locking, browser locking, } and \emph{Master Boot Record (MBR) locking}.

\emph{Screen Locking} 
ransomware 
lock the system's 
graphical user interface and prevent 
access while demanding a ransom to lift the restriction. They can lock the screen of the victim using different methods, including 
employing OS functions (e.g., CreateDesktop) to create a new desktop and making it persistent~\cite{cuttingGordian}. Some ransomware families like Reveton~\cite{Cimpanu_2019} can download images or HTML pages from C\&C servers, and create their lock banner dynamically. Screen locking ransomware 
can also target mobile devices. In this respect, screen locking is frequently applied by Android ransomware families~\cite{Snow_2016}. To lock the mobile device, 
while some families like LockerPin set the specific parameters to Android System APIs to make Android screen persistent, others 
like WipeLocker disable the specific buttons (e.g., Home Button) of mobile devices~\cite{DNADROID}. 

\emph{Browser Locking} ransomware families lock web browser of the victim and demand a ransom. Attackers lock browsers of victims by redirecting victims to a web page that contains a malicious JavaScript code. Unlike other malicious ransomware tactics, recovery from  browser lockers is relatively simpler. To scare victims, such ransomware can display a ransom message stating that the computer has been blocked due to violation of law. 

\emph{MBR Locking} ransomware families, such as Seftad~\cite{seftad1}, target Master Boot Records (MBR) of the system. MBR of a system contains the required information 
to boot the operating system. So, the result of such a malicious action aims to prevent the system from loading the boot code 
either by replacing the original MBR with a bogus MBR, or by encrypting the original MBR. 

\subsubsection{Data Exfiltration}
In addition to encryption and destruction, some ransomware families, especially the 
recent ones, also try to steal victim's valuable information (e.g., credit card information, corporate documents, personal files, etc.)~\cite{stealData}. In fact, a few ransomware families demand two ransom payments. As such, one of the payments to send the key to decrypt the files, and the other one to prevent publishing the stolen information~\cite{tworansom}. The motivation of such actions is to demand more ransom amounts from the victims and to speed up the payment process.

\subsection{Classification by Extortion Method}
The main objective of ransomware is extorting money (i.e., ransom payment) from  victims. The fundamental characteristic of ransomware extortion methods is \emph{anonymity}. Throughout the evolution of ransomware, cybercriminals utilized different extortion methods. Payment methods such as premium-rate text messages, pre-paid vouchers like Paysafe card have been utilized by ransomware families. However, cryptocurrencies such as Bitcoin are the most preferred method to extort money at the moment due to their decentralized and unregulated nature, pseudo-anonymity, and not being subject to local law authorities.

\subsection{Taxonomy of Notable Ransomware Families}\label{sec:taxonomy-notable-ransomware}
\begin{figure*}[htbp]
\centering
\subfigure[Infection Methods]{%
\includegraphics[scale= 0.32]{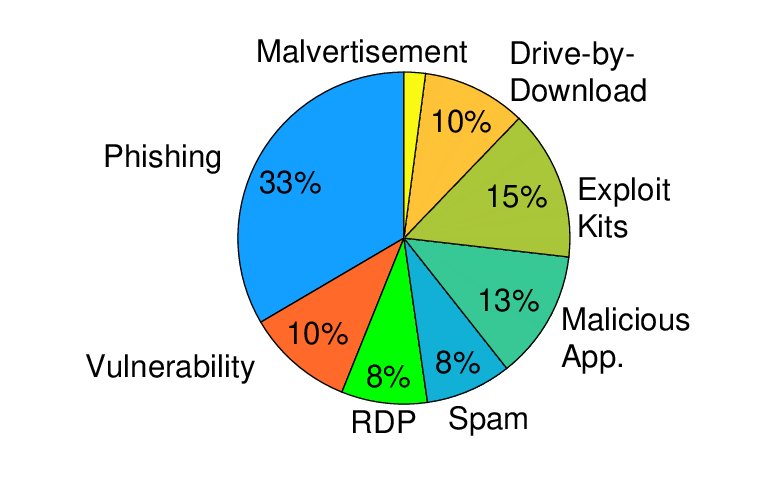}}
\subfigure[C\&C Communication]{%
\includegraphics[scale= 0.32]{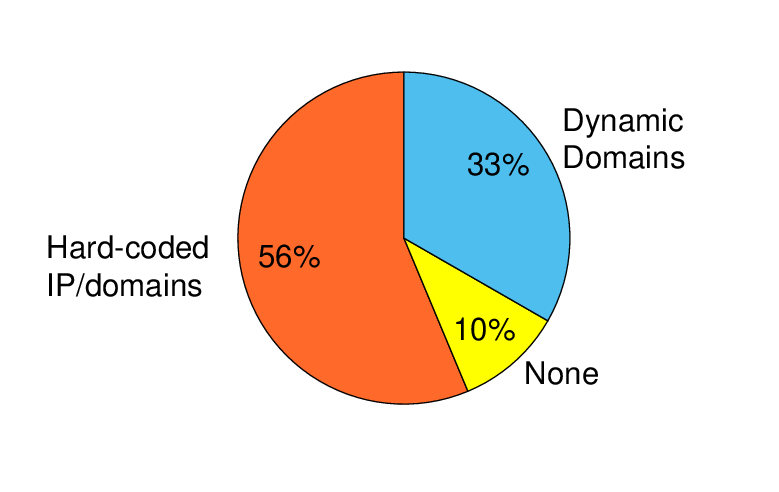}}
\subfigure[\harun{Target Platforms}]{%
\includegraphics[scale= 0.32]{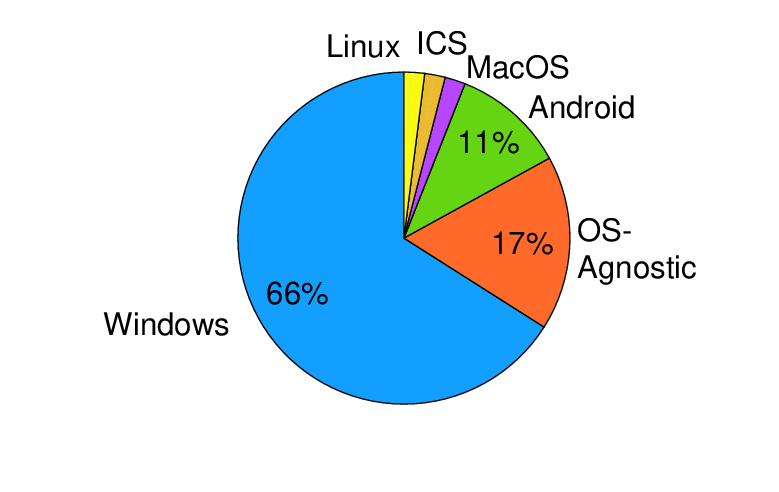}}
\subfigure[Encryption Techniques]{%
\includegraphics[scale= 0.32]{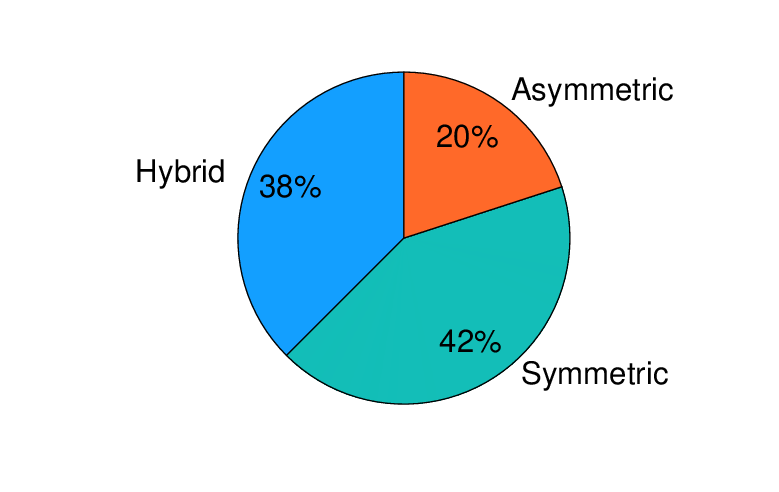}}
\subfigure[Destruction Methods]{%
\includegraphics[scale= 0.32]{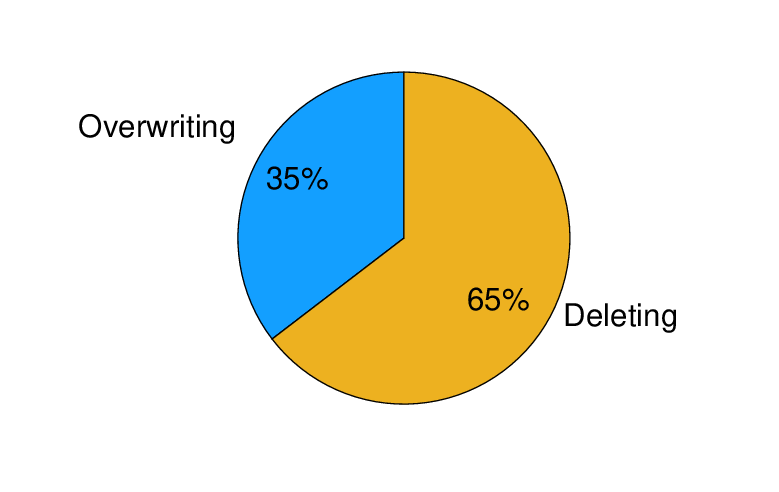}}

\vspace{-0.5em}
\caption{Distribution of infection, C\&C communication, \harun{target platforms}, encryption techniques, and destruction methods of the notable ransomware families from 1989 to 2020.}
\label{fig:ransomware-tax-graph}
\vspace{-1.5em}
\end{figure*}

Table~\ref{tab:ransomware-strains} outlines the taxonomy of notable ransomware families that were observed in the wild between 1989-2020. \harun{We used major attack instances~\cite{seftad1,Infostealer_2020,Cimpanu_2019,bates1990trojan,bates1990trojan2}, and academic papers~\cite{berrueta:2020,dargahi:2019,maigida:2019,cuttingGordian,wannaCryAnalysis} as sources to build Table~\ref{tab:ransomware-strains} in our work. We also used popular blog posts~\cite{TheriseofAndroid,Locky,Cannell_2013,tworansom,wannacry,Archiveus,symantecEvolution,randsom} of the security companies not to miss any relevant ransomware family.} 
As some ransomware families evolved in time to enhance their abilities and/or evade defense mechanisms, multiple variants for such ransomware emerged. We considered only the first version of such ransomware in the table. Based on the taxonomy of notable ransomware families in Table~\ref{tab:ransomware-strains}, Figure~\ref{fig:ransomware-tax-graph} shows the distribution of C\&C communication, destruction behavior, \harun{target platforms}, encryption technique, and infection methods. 


Although ransomware is very different from other malware types, ransomware strains utilized traditional malware infection techniques to infect victims. Our findings remark that ransomware most commonly infects a victim system via phishing e-mails. As shown in Figure~\ref{fig:ransomware-tax-graph}, $33\%$ of the notable ransomware families use phishing for infection. Exploit kits, drive-by-download, and malicious applications are the other popular infection methods used by ransomware. 


As outlined in Table~\ref{tab:ransomware-strains}, majority of the notable ransomware families employ some sort of communication either using IP addresses or domains hard-coded into ransomware binaries, or domain names dynamically generated via DGAs. As shown in Figure~\ref{fig:ransomware-tax-graph}, more than $50\%$ of the families use hardcoded IP addresses or domains, while a very small portion of the families use dynamically generated domains. We also see that almost one third of the families do not employ any C\&C communication as shown in Figure~\ref{fig:ransomware-tax-graph}. 

\harun{Considering the target platforms, Windows is the most popular target of the notable ransomware strains. OS-agnostic ransomware and Android-based ransomware follow Windows-based ransomware in popularity. As Windows is the most popular operating system used by end-users, organizations, and enterprises, notable ransomware families target Windows platforms to widen their victims and increase their profits. OS-agnostic programming languages such as Python play an important role for ransomware authors to develop platform-independent ransomware strains and the ratio for OS-agnostic ransomware may increase in the following years. }

\begin{table*}
    \caption{{Summary of Notable Ransomware Families from 1989 to 2020.}}\label{tab:ransomware-strains}
    \centering
    \begin{adjustbox}{width=\textwidth,totalheight=\textheight,keepaspectratio}
    \begin{tabular}{|c|c|c|c|c|c|c|c|c|c|c|c|c|c|}
    \hline
        \multirow{2}{*}{\textbf{Family}}&\multirow{2}{*}{\textbf{First Seen}}& \multirow{2}{*}{\textbf{Infection}} & \multirow{2}{*}{\textbf{Platform}} &\multicolumn{7}{c|}{\textbf{Characteristics}}   \tabularnewline

\cline{5-11} 
&   & & &  \textbf{C\&C Comm.} & \textbf{Encryption}  &\textbf{Destruction } &  \textbf{Exfiltration} & \textbf{Locking} & \textbf{Del. Shadow} & \textbf{Extortion}  \tabularnewline
\hline 

PC CYBORG& 1989  & Phishing & Windows &  None & Custom & Overwrite  &  & & &  Cash \tabularnewline
\hline
GPCode& 2008 & Drive-by-download & Windows & None  & RSA &  Overwrite &  & & &  Prepaid Voucher  \tabularnewline
\hline

Archiveus& 2008 & Drive-by-download & Windows & None & RSA & Delete &  & & &  Prepaid Voucher \tabularnewline
\hline

Randsom.C& 2008  & Spam & Windows &  None & None  & None  &  \checkmark  &  &  & SMS  \tabularnewline
\hline

Seftad& 2008  & Spam & Windows &  None &  MBR & Overwrite   &  &\checkmark  &  &Prepaid Voucher \tabularnewline
\hline

Krotten & 2008  & Spam & Windows &  None &  RSA & Overwrite   &  &    &  &Prepaid Voucher \tabularnewline
\hline

Urausy& 2009  & Phishing & Windows &  None &None  & None  &   &  \checkmark  &  & Prepaid Voucher  \tabularnewline
\hline

Winlock & 2010  & Spam & Windows &  Hard-coded & None  & None  &  \checkmark  &  &\checkmark  & SMS  \tabularnewline
\hline

Reveton & 2012  & Phishing & OS-Agnostic & None  & None  &   None &  \checkmark & \checkmark &  & Prepaid Voucher\tabularnewline
\hline

CryptoLocker & 2013  & Phishing & Windows &  Hard-coded & RSA &  Overwrite &  \checkmark &   & \checkmark & Bitcoin  \tabularnewline
\hline

CryptoWall& 2013  &  Phishing & Windows &  Hard-coded & RSA  &  Overwrite  &  &  & \checkmark & Prepaid Voucher \tabularnewline
\hline

FakeDefender & 2013  & Malicious App. & Android & None &  None &  None  &  & \checkmark &  & Prepaid Vouchers  \tabularnewline
\hline

Lockdroid & 2014  & Malicious App. & Android & None &  None &  None  &  & \checkmark &  & Prepaid Vouchers  \tabularnewline
\hline

SimpLocker & 2014  & Malicious App. & Android &  Hard-coded & AES  &  Overwrite  &  &  &  & Bitcoin \tabularnewline
\hline

TorrentLocker & 2014  &  Phishing & Windows & Dynamic Domains & AES+RSA  & Delete &  &  & \checkmark & Prepaid Voucher  \tabularnewline
\hline

TrolDesh& 2014  & Phishing & Windows &   Hard-coded & AES  &  Delete &  & & \checkmark & Bitcoin  \tabularnewline
\hline

CryptoDefense& 2014  &  Phishing & Windows &  None &  RSA  &  Delete &  &  & \checkmark & Bitcoin  \tabularnewline
\hline

CTBLocker & 2015  &  Phishing & Windows &  None & ECC   &  Delete   &  & & \checkmark & Bitcoin  \tabularnewline
\hline

TeslaCrypt & 2015  & Exploit kits & Windows &  Hard-coded &  AES  &  Delete   &  &   &  & Bitcoin  \tabularnewline
\hline

Fusob & 2015  & Phishing & Android &  None &  None &  None  &  &  \checkmark &  & Giftcards \tabularnewline
\hline

Chimera& 2015   &  Phishing & Windows &   Hard-coded &  AES  &  Delete &  &  & \checkmark & Bitcoin  \tabularnewline
\hline
LinuxEncoder& 2015  &  Vulnerability & Linux &   Hard-coded  & AES+RSA   &  Delete &  & &  & Bitcoin  \tabularnewline
\hline

Ransom32& 2016  &  Malvertisement  & OS-Agnostic &  Hard-coded &  AES & Overwrite  &  & &  \checkmark& Bitcoin  \tabularnewline
\hline

Dharma& 2016  & RDP & Windows &  Hard-coded &  AES & Delete  &  &  &  \checkmark& Bitcoin  \tabularnewline
\hline

Locky & 2016  &  Phishing & Windows &  Dynamic Domains &  AES+RSA  &  Delete  &  &   & \checkmark & Bitcoin  \tabularnewline
\hline

Cerber& 2016  & Phishing & Windows &  None & AES   &  Overwrite  &  & &  & Bitcoin  \tabularnewline
\hline

Jigsaw& 2016  & Phishing & Windows &   Hard-coded & AES+RSA   &  Delete &  & & \checkmark  & Bitcoin  \tabularnewline
\hline

KeRanger& 2016 & Malicious App. & macOS &  Hard-coded & AES+RSA   &  Delete  &  &  & \checkmark & Bitcoin  \tabularnewline
\hline

Petya & 2016  & Exploit kits &Windows &  Hard-coded & AES & Overwrite &  & \checkmark&  & Bitcoin  \tabularnewline
\hline

DMALocker& 2016  & Exploit kits  & OS-Agnostic &  Hard-coded & AES & Delete &  & & \checkmark & Bitcoin  \tabularnewline
\hline

Sage& 2017  &  Drive-by-download & Windows &   Hard-coded& AES+RSA   &  Delete  &  &  &  & Bitcoin  \tabularnewline
\hline

BadRabbit& 2017  &  Drive-by-download & Windows &  Dynamic Domains  & AES+RSA   &  Delete  &  &  &  & Bitcoin  \tabularnewline
\hline
WannaCry & 2017  &  Vulnerability  & Windows &  Hard-coded&  AES+RSA &  Delete  &  &  &  & Bitcoin  \tabularnewline
\hline

GoldenEye& 2017  & Exploit kits & Windows &  Hard-coded& AES  &  Overwrite &  & & \checkmark & Bitcoin  \tabularnewline
\hline

SamSam& 2018  &   RDP & Windows &   Hard-coded &  RSA &  Delete &  &  & \checkmark & Bitcoin  \tabularnewline
\hline

GandCrab&2018 &  Drive-by-download  & OS-Agnostic &  Hard-coded & AES &  Overwrite &  & & \checkmark & Bitcoin  \tabularnewline
\hline

Sodikonibi& 2019  & Exploit kits & Windows &  Hard-coded & AES+ECC &  Overwrite &  &\checkmark & \checkmark & Bitcoin  \tabularnewline
\hline

Robbinhood & 2019  & RDP & Windows &  Hard-coded & AES+RSA &  Delete&   \checkmark & & \checkmark & Bitcoin  \tabularnewline
\hline
Maze& 2019 & Exploit kits & OS-Agnostic & Hard-coded &  AES+RSA & Overwrite  &  &  &  \checkmark& Bitcoin  \tabularnewline
\hline
Ryuk& 2019  & RDP & OS-Agnostic &  Hard-coded &  AES+RSA & Overwrite  &  &  &  \checkmark& Bitcoin  \tabularnewline
\hline

MegaCortex& 2019  & Exploit-kits &OS-Agnostic &   Hard-coded& AES &  Overwrite &  &  & \checkmark & Bitcoin  \tabularnewline
\hline
LockerGaga & 2019  & Phishing & Windows &  None & AES+RSA  &  Delete &  & & \checkmark & Bitcoin  \tabularnewline
\hline
Ekans& 2019  & Vulnerability & ICS  &  Hard-coded & AES+RSA  &  Delete &  & & \checkmark & Bitcoin  \tabularnewline
\hline

PureLocker& 2020  & Vulnerability & OS-Agnostic & Dynamic Domains  &  AES  & Delete  &    &  &\checkmark  & Proton\tabularnewline
\hline

Tycoon& 2020  & Vulnerability & Windows & Dynamic Domains  &  AES  & Delete  &    &  &\checkmark  & Proton\tabularnewline
\hline

CovidLock& 2020  & Malicious App. & Android & None & None & None &    &  \checkmark & & Bitcoin\tabularnewline
\hline

Corona& 2020  & Phishing & Windows & None &  AES+RSA &  Overwrite  & \checkmark &    & \checkmark & Bitcoin 
\tabularnewline
\hline
    \end{tabular}
    \end{adjustbox}
    \end{table*}

In terms of the malicious action, we can see that majority of the notable ransomware families are cryptographic ransomware using various encryption techniques. As Table~\ref{tab:ransomware-strains} shows, compared to cryptographic ransomware, locker ransomware variants were more dominant between 2009-2013. This may be due to cryptographic ransomware families embedding their keys to the binaries in the early years lack of built-in cryptographic frameworks and proper key management strategies. However, after the emergence of perfectly developed potent cryptographic strains, cybercriminals almost entirely started developing cryptographic families for PCs. Nevertheless, as we discussed before, locker ransomware families remain a crucial threat to Android-based platforms.

The ransomware's main purpose is to perform a malicious activity (i.e., encryption or locking) on the victim system and demand a payment. Given its effectiveness, cryptographic ransomware still preserves its intensity. 
As shown in Figure~\ref{fig:ransomware-tax-graph}, symmetric-key encryption is more popular than both asymmetric-key encryption and hybrid encryption techniques. We can see that only $20\%$ of the notable ransomware families solely employed asymmetric-key encryption. Although the percentage of hybrid encryption seems to be lower than the symmetric-key encryption, the use of hybrid encryption has gradually increased in the recent years as outlined on Table~\ref{tab:ransomware-strains}. In terms of the specific encryption algorithm, even though some families have utilized different algorithms, ECC, AES and RSA are the most common ones.

To destroy the victim's original data, cryptographic ransomware families either delete or overwrite the original file. 
As shown in Figure~\ref{fig:ransomware-tax-graph}, almost two third of the notable ransomware families delete the original files after the encryption process, and only one third of the families prefer overwriting. In addition, as Table~\ref{tab:ransomware-strains} shows, deletion of the shadow copies to  
prevent the recovery of the victim files is very common among families.



As Table~\ref{tab:ransomware-strains} outlines, cryptographic ransomware strains mostly use cryptocurrencies as their payment method. This is different for mobile platforms that mobile ransomware strains mostly use SMS payments to obtain their ransom. Differently, the strains that launch targeted attacks to the enterprises require victims to contact via encrypted e-mail services such as Proton \cite{encryptedE-mail} to negotiate the ransom amount and payment process.

\section{Ransomware Defense Research}\label{sec:taxonomy-research}

\begin{figure*}[htbp]
\centering
\includegraphics[width=\linewidth]{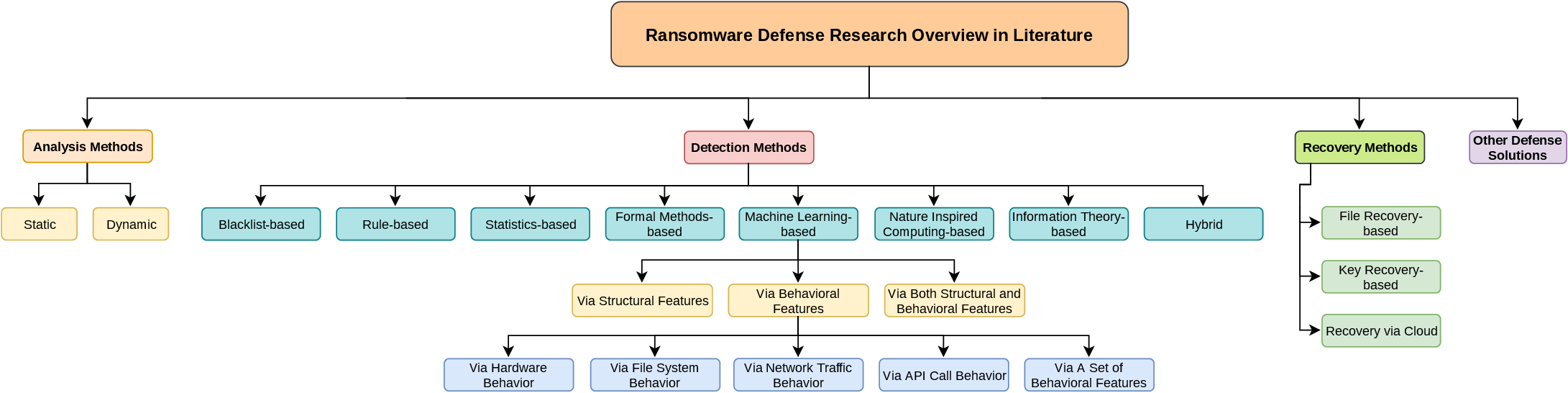}
\caption{An overview of ransomware defense research \emph{in literature}.}
\label{fig:ransomwareResearch}
\end{figure*}
 
In this section, we give an extensive overview of ransomware defense research. As shown in Figure~\ref{fig:ransomwareResearch}, ransomware defense research can be divided into four categories: \emph{analysis, detection, recovery}, and \emph{other defense research}. In this survey, we provide a taxonomy of each research domain with respect to target platforms of \emph{PCs/workstations, Mobile Devices,} and \emph{IoT/CPS}. Based on the target platforms, we firstly give an overview of various ransomware analysis techniques, then categorize and explain ransomware detection systems, and finally summarize the recovery mechanisms. In addition to these three categories, there exist some studies that do not fall into any of the aforementioned categories that were summarized under \emph{Other Methods} category in this survey.

\subsection{Ransomware Analysis Research}\label{sec:taxonomy-analysis}
Ransomware analysis includes activities to understand the behavior and/or characteristics of ransomware. Similar to traditional malware analysis, ransomware analysis techniques can be categorized as \emph{static} and \emph{dynamic}. 

\emph{Static analysis} aims to understand whether a sample is a ransomware or not by extracting structural information from the sample without actually running it. To analyze a sample without running it and still obtain useful information, researchers disassemble sample binaries and extract information regarding the structure/content of the sample. Static analysis is usually fast and safe since the sample is not run. However, malware authors employ concealment (i.e., obfuscation, polymorphism, encryption) and anti-disassembler techniques to make the static analysis efforts harder, and evade the defense schemes that use the structural features obtained via static analysis.

\emph{Dynamic analysis} of ransomware consists of running the sample and observing the behavior to determine if the sample is a ransomware or not. Dynamic analysis is performed via running the samples inside an isolated environment (i.e., sandbox) to avoid a possible damage caused by the analyzed sample. Researchers can use hooking techniques and functionalities provided by the sandbox environment to monitor the behavior of the sample. Since it requires an isolated environment and actual activation of ransomware, it is costly in terms of time and resources compared to static analysis. Concealment techniques and anti-disassembler techniques effective against static analysis cannot be effective against dynamic analysis since those approaches cannot conceal the behavior of the ransomware. However, ransomware authors utilize anti-debugging techniques, sandbox fingerprinting approaches, and logic bomb schemes (e.g., activating the malicious behavior based on a certain time or event happening) to make dynamic analysis efforts harder.

Static and dynamic analysis have their own advantages and disadvantages, which result in researchers to use both of the approaches in \emph{hybrid} analysis. In this section, we categorize and give overview of static and dynamic analysis features extracted in ransomware research.

\subsubsection{Ransomware Analysis in PCs/workstations}\hfill 

In this subsection, we give an overview of structural and behavioral features obtained via static and dynamic analysis of ransomware samples targeting PCs/workstations respectively.
\vspace{0.4em}

\textbf{\emph{Structural features}} obtained from ransomware for PCs/workstations consist of \emph{file hashes, header information, function/API/system calls, strings, opcodes}, and \emph{file types}. Researchers obtain these features from ransomware samples targeting PCs/workstations without running the samples.

\vspace{0.2em}
\noindent\emph{Strings:} 
Ransomware displays a ransom note at the end of the destruction process. In addition, ransomware binaries include strings such as \emph{encrypt}, \emph{bitcoin}, specific \emph{IP addresses}~\cite{berrueta:2020}. Those strings that are obtained from samples can be signs of ransomware.

\vspace{0.2em}
\noindent\emph{File Hashes:}  
Hash digest of a sample can be looked-up against a database of known ransomware hashes to detect ransomware. However, defense mechanisms relying only on the hash values can be easily evaded by adversaries applying small manipulations on the ransomware.

\vspace{0.2em}
\noindent\emph{Header Information:} Headers of samples (e.g., Portable Executable (PE) header in Windows, Executable and Linkable Format (ELF) headers in Linux, and Mach-O headers in macOS) can give valuable information regarding the malicious characteristics of a sample. Researchers can analyze section information, symbols, optional headers, etc. by checking the header of a sample.  

\vspace{0.2em}
\noindent\emph{Function/API/System Calls:} Functions/system/API calls can be obtained via static analysis. These calls can be used by applications for crucial operations such as encryption, memory management, file system, or network operations that may discriminate ransomware from benign applications~\cite{miningAPICall}.

\vspace{0.2em}
\noindent\emph{Opcodes:} Instruction opcodes and patterns of opcode sequences can be used to determine if a sample is ransomware or not.
\vspace{0.4em}

\textbf{\emph{Behavioral features}} obtained from ransomware for PCs/workstations include \emph{registry activity, host logs, process activity, file system activity, inputs and outputs of function/API/system calls, I/O access patterns, network activity, resource usage,} and \emph{sensor readings}. Researchers obtain these features from ransomware samples targeting PCs/workstations via running them in analysis environments.

\vspace{0.2em}
\noindent\emph{Registry Activity:} During the installation process in Windows platforms, ransomware performs changes in the registry to remain persistent after system reboots~\cite{practicalMalware}. \harun{However, not only ransomware but also other malware perform similar changes in the registry to be persistent. Therefore, registry activity can be utilized as an additional feature to detect ransomware.}

\vspace{0.2em}
\noindent\emph{Host Logs:} Extracted events from the host logs can be used to capture ransomware actions in the system~\cite{patternExtraction}. 

\vspace{0.2em}
\noindent\emph{File System Activity:} Ransomware scans the file system, encrypts all or a subset of files, deletes or overwrites the existing files. Therefore, file system activity can be used for ransomware detection.

\vspace{0.2em}
\noindent\emph{Function/API/System Calls:} While function/API/system calls that can be made by a sample can be obtained via static analysis, the actual calls made, parameters, results, and sequences can be monitored via dynamic analysis.

\vspace{0.2em}
\noindent\emph{I/O Accesses:} The operations performed by ransomware (i.e., encryption, deletion or overwrite) involve repetitive I/O access activities of read, write, and delete. Therefore, patterns of I/O access can be used to detect ransomware~\cite{cuttingGordian}.

\vspace{0.2em}
\noindent\emph{Network Activity:} Communication-related features such as source and destination IP addresses, ports, domain names, and protocols can be used by researchers to determine if a sample displays ransomware-like communication behavior.

\vspace{0.2em}
\noindent\emph{Resource Usage:} 
Since ransomware relies on encryption operation, high \emph{CPU usage} or \emph{memory usage} can be a sign for the existence of ransomware in the system~\cite{deception}.

\vspace{0.2em}
\noindent\emph{Sensor Readings:} On-board sensor readings of PCs/workstations can give a clue on the abnormal activity which can signify the existence of ransomware in the system~\cite{Taylor2017SensorbasedRD}.

\subsubsection{Ransomware Analysis in Mobile Devices}\hfill

In this subsection, we give an overview of structural and behavioral features obtained from static and dynamic analysis of ransomware samples targeting mobile devices respectively. 

\vspace{0.4em}
\textbf{\emph{Structural features}} obtained from ransomware for mobile devices are \emph{strings, opcodes, application images, permissions requests} and \emph{API packages}.

\vspace{0.2em}
\noindent\emph{Strings:} The strings that are extracted from the packaged mobile application can be used as a feature to detect mobile ransomware. Such strings can contain IP addresses, domain names, ransom notes, etc., which can be helpful to detect ransomware.

\vspace{0.2em}
\noindent\emph{Opcodes:} Instruction opcodes that are obtained from the disassembled application byte-code can be used to understand if a mobile application has the characteristics of ransomware. 

\vspace{0.2em}
\noindent\emph{Application Images:} Extracted images from the application may contain ransom related material (i.e., ransom message image)~\cite{DNADROID}, and thus be used as a feature to detect mobile ransomware.

\vspace{0.2em}
\noindent\emph{Permissions:} Mobile applications require permissions to be approved by the users to access and utilize resources of the mobile device. Permissions can be an indicator of ransomware intention of a mobile application.

\vspace{0.2em}
\noindent\emph{API Packages:} API packages can be extracted from the source code of a mobile application to determine the malicious encryption or locking characteristics.

\vspace{0.4em}

\textbf{\emph{Behavioral features}}  obtained from ransomware for mobile devices are \emph{function/API/system calls, user interaction, file system features,} and \emph{resource usage}.

\vspace{0.2em}
\noindent\emph{Function/API/System Calls:} Researchers can detect mobile ransomware variants by analyzing the function/API/system calls made by a mobile application while running.

\vspace{0.2em}
\noindent\emph{User Interaction:} Matching the user's interactions with the events taking place while the application is running can be used to detect the presence of a ransomware. 

\vspace{0.2em}
\noindent\emph{File System Features :} Like in PCs/workstations, the features extracted from file system of a mobile device can be used to understand the presence of ransomware.

\vspace{0.2em}
\noindent\emph{Resource Usage:} Similar to PCs/workstations, abnormalities in the resource usage patterns on a mobile device, such as power consumption can be a sign of the presence of a mobile ransomware.

\subsubsection{Ransomware Analysis in IoT/CPS Platforms}\hfill 

In this section, we give an overview of structural and behavioral features extracted from ransomware that can target IoT/CPS platforms. Since ransomware defense research for IoT/CPS environments is in its infancy at the moment, only a few studies exist in the literature. Considering the existing ransomware defense research targeting IoT/CPS platforms, only behavioral features, namely, \emph{network activities} were used in the literature.

\noindent\emph{Network Activity:} Network-related features are captured by researchers within the IoT/CPS environment to find out the communication patterns signifying the presence of ransomware~\cite{deepIoT}.

\subsection{Ransomware Detection Research} \label{detection-research}
In this subsection, we categorize and summarize existing detection mechanisms for ransomware with respect to target platforms. Based on the employed methodology, we categorize detection systems into eight categories: 
\begin{itemize}
\item \emph{\textbf{Blacklist-based:}} the system detects ransomware using a list of malicious domain names or IP addresses that are known to be used by ransomware families.
\item \emph{\textbf{Rule-based:}} the system detects ransomware using rules that are constructed using the analysis features. Rules can be either the rules compatible with malware detection engines (e.g., YARA), maliciousness scores, or threshold values.
\item \emph{\textbf{Statistics-based:}} the system detects ransomware using statistics on features indicating that the sample is a ransomware. 
\item \emph{\textbf{Formal Methods-based:}} the system detects ransomware using a formal model that can discriminate malicious and benign patterns. 
\item \emph{\textbf{Nature Inspired Computing-based:}} the system detects ransomware using techniques inspired from the nature and biology. 
\item \emph{\textbf{Information Theory-based:}} the system detects ransomware using information theory approaches (e.g., entropy). Encryption operation performed by cryptographic ransomware strains results in changes in the information content of the files. For this reason, significant changes in entropy is considered as an indicator of ransomware by several researchers. However, benign encryption, compression, and file conversion operations on already compressed file formats also result in high entropy values. Therefore, entropy is mostly used as a supportive feature for ransomware detection.
\item \emph{\textbf{Machine Learning-based:}} the system detects ransomware via ML models that are built using a set of analysis features. ML-based ransomware detection systems use either structural features, behavioral features, or both. Structural features are obtained by researchers via static analysis of ransomware binaries. By using the structural features in the training process of ML classifiers, detection systems can detect the patterns in ransomware binary structures. Behavioral features on the other hand are obtained via dynamic analysis of ransomware binaries. By using behavioral features in the training process of ML classifiers, detection systems can detect the patterns in the behavior of ransomware binaries.
\item \emph{\textbf{Hybrid:}} the system detects ransomware via a set of the detection techniques.
\end{itemize}

\subsubsection{Ransomware Detection for PCs/Workstations}\hfill

In this subsection, we provide an overview of rule-based, machine learning-based, deep learning-based, information theory-based and other ransomware detection systems for PCs/workstations. 

\textbf{\emph{Blacklist-Based Detection}}\hfill

Akbanov et al.~\cite{akbanov:2019} examined the behavior of WannaCry ransomware on SDN, and proposed an SDN-based ransomware detection method. Their detection system runs as an application on the SDN controller and monitors the network traffic for the appearance of malicious domain names or the IP addresses used by WannaCry. Once a matching flow is detected, rules to block that malicious traffic are generated.

\textbf{\emph{Rule-Based Detection}}\hfill

YARA rules are created by the rule-based ransomware detection system of Medhat et al.~\cite{medhatetal} using API calls of file and cryptography libraries, strings, and file extensions from ransomware binaries. Using the YARA scanner, their system scans each sample, and assigns a score based on the existence of these features in the samples.

Maliciousness scores are calculated in CryptoDrop~\cite{cryptoDrop} and REDEMPTION~\cite{redemption} to detect ransomware. While file type changes, similarity and entropy of files, deletion of files, and file type funnelling are employed by CryptoDrop~\cite{cryptoDrop} to determine the score, REDEMPTION~\cite{redemption} utilizes directory traversal, file type change, access frequency, and file content features (i.e., entropy ratio of data blocks, file content overwrite, delete operation) for the score calculation. In Amoeba~\cite{minetal} proposed by Min et al., the risk indicator for ransomware attack is calculated for every write operation on SSD. Amoeba uses intensity (number of write requests), similarity (similarity of old and new data), and entropy of page write operations to compute the risk indicator and detect ransomware. In UNVEIL~\cite{EnginKirda}, a ransomware analysis system that generates an artificial user environment is developed which monitors file-access patterns and the buffer entropy. In addition, UNVEIL detects locker ransomware by investigating ransom notes by taking screenshots of the analysis environment, and checking if structural similarity of the screenshots are above a threshold.

 
In terms of the rule-based systems that use network traffic features, REDFISH~\cite{redFish} was proposed to detect ransomware that encrypt files in the network shared volumes. It monitors the traffic between PCs/workstations and network shared volumes, and applies three threshold values on number of files deleted, time interval between deletion events, and average R/W speed. In the work of Cabaj et al.~\cite{Cabaj:2016}, centroids were built for the HTTP POST message content sizes of ransomware families. Ransomware is detected if Euclidean distance of three consecutive HTTP POST message content sizes from the centroids are below a threshold value.


\textbf{\emph{Statistics-Based Detection}}\hfill

Palisse et al. proposed a statistics-based ransomware detection system, namely Data Aware Defense (DAD)~\cite{DaD}. DAD focuses on features obtained from write operations such as buffer content, size, offset, file name, process id and name, and thread id. Considering the last 50 write operations, it uses the chi-square goodness-of-fit test and checks whether the obtained median value is above a certain threshold.

\textbf{\emph{Information Theory-Based Detection}}\hfill

Since benign encryption, compression, and file conversion operations on already compressed file formats also result in high entropy values, several researchers~\cite{cryptoDrop,EnginKirda,redemption,minetal,RansomBlocker,shieldFS} used entropy as a supportive feature for their detection systems. However, there exist a few studies which used entropy as the primary feature to detect ransomware. In this regard, Lee et al.~\cite{leeetal2} proposed a detection system which aims to detect ransomware and also prevent ransomware affecting the cloud storage backups. Their system calculates the entropy of the files that are about to be transferred to the cloud storage systems and compares it to a threshold value to detect ransomware.

\textbf{\emph{Formal Methods-Based Detection}}\hfill

In \cite{databasedetect}, Iffländer et al. proposed DIMAQS (Dynamic Identification of Malicious Query Sequences) for detection of ransomware targeting database servers. DIMAQS utilizes colored Petri nets-based classifier to detect the malicious query sequences made by ransomware to target database servers. 

\textbf{\emph{Nature Inspired Computing-Based Detection}}\hfill

An Artificial Immune System-based ransomware detection system was proposed by Lu et al.~\cite{lu:2020}. The proposed system uses API call n-grams as antigens and employs a double-layer negative selection algorithm to discriminate ransomware from benign applications.

\textbf{\emph{Machine Learning-Based Detection}} \hfill

\noindent\textbf{\emph{Via Structural Features:}}
In terms of the ML-based ransomware detection systems for PCs/workstations using structural features, researchers employed instruction opcodes, API calls, and DLLs. 

Instruction opcode sequences of binaries were used by \cite{Xiao:2018, salehetal, SVMbaldwin, ZHANG2020708} to build ML classifiers for ransomware detection. Opcode n-grams were used by Zhang et al.~\cite{ZHANG2020708} to build a Deep Neural Network (DNN)-based classifier and by Xiao et al~\cite{Xiao:2018} to build various ML classifiers. While opcodes of various instructions (i.e., data process, arithmetic, logic, and control flow) were used to build a Hidden Markov Model (HMM) by Saleh et al.~\cite{salehetal}, opcode densities were used by Baldwin et al.~\cite{SVMbaldwin} to build a Support Vector Machine (SVM) classifier for ransomware detection.

API call frequency was used by Martinelli et al.~\cite{Martinelli2018PhylogeneticAF} for ransomware detection. They extracted API calls from ransomware samples via static analysis, and trained a Random Forest (RF) classifier with API call frequencies to detect ransomware. 

Instead of using a single structural feature, Poudyal et al.~\cite{poudyaletal} employed multiple features in which they extracted opcodes and DLLs of binaries, and built an RF classifier. 

\vspace{0.4em}
\noindent\textbf{\emph{Via Behavioral Features:}}
In terms of the ML-based ransomware detection systems proposed for PCs/workstations using behavioral features, researchers monitored and/or analyzed hardware, file system, network traffic, and API call behaviors.

\vspace{0.2em}
\textbf{\emph{Via Hardware Behavior:}}
PC/workstation hardware including storage hardware, on-board sensors, and memory dumps were monitored by researchers for ransomware detection. 
 
I/O operations performed by CPU on storage devices were used by researchers for ransomware detection. However, monitoring of I/O operations and storage hardware results in high granular data (e.g., block address, read/write type, size of data) which makes detection harder since higher level data such as process and file information cannot be obtained by I/O operations monitoring~\cite{SSD_Insider}. Baek et al.~\cite{SSD_Insider} proposed SSD-Insider, which monitors I/O request headers to detect ransomware-like patterns in overwriting actions on the SSD. They trained a Decision Tree (DT) classifier with six overwriting-related features obtained from I/O request headers. In RansomBlocker~\cite{RansomBlocker}, Park et al. introduced an encryption-aware ransomware protection system that examines entropy of the data written to the host SSD. Their system uses a Convolutional Neural Network (CNN)-based classifier to discriminate high entropy benign write operations from encrypted write operations.

Cohen and Nissim~\cite{COHEN2018158} utilized Volatility framework to monitor the volatile memory of a virtual machine. They extracted DLL and process features, kernel modules and callbacks, privileges, services, handles, etc. from the memory dumps, and trained various ML models to detect ransomware in private clouds. Taylor et al.~\cite{Taylor2017SensorbasedRD} leveraged hardware sensor monitoring to detect ransomware behavior by observing its possible side-channel effects on the PC hardware. They used the readings of 59 different on-board sensors, and trained a Logistic Regression ML model.
\harun{The work presented in \cite{IntelDetect} employed a CPU-based behavioral monitoring approach to detect ransomware in Intel vPro platform-based PCs. They utilized CPU level telemetry and ML heuristics to detect the encryption operation of ransomware and possibly other malware in the hardware level. }


\vspace{0.2em}
\textbf{\emph{Via File System Behavior:}}
Instead of monitoring the hardware, some researchers aimed to detect ransomware at a higher level via monitoring file system activities. Compared to hardware behavior, file system behavior monitoring can provide a lower granular data allowing to obtain file and process information. Several researchers~\cite{KnowAbnormal.,Goyal:2020,RWGuard,patternExtraction,HSRAutomated,ransHaunt,twoStageML,zuhairetal,Egunjobietal,migidaetal,eldeRan,Abbasietal,Ashrafetal,Jethvaetal} used file system behavior features with other structural or behavior features. However, there exist a few studies which used file system behavior as the primary source to detect ransomware. Continella et al.~\cite{shieldFS} proposed ShieldFS that detects ransomware by capturing short-term and long-term file system activity patterns. They trained RF classifiers such that each classifier is trained on the file-system activity features on different time scales. They used number of files accessed, read, renamed, moved, or written, entropy of write operations, and folder listing operations as discriminating features for ransomware detection.

\vspace{0.2em}
\textbf{\emph{Via Network Traffic Behavior:}}
Since ransomware usually communicates with its C\&C server for key exchange or data exfiltration, some researchers aimed to detect ransomware in the networked-devices by observing the network traffic. The monitoring schemes monitor either the traffic of the host, or the traffic of the complete network, or subnet it is deployed to.

In terms of the host-based traffic monitoring, the works~\cite{Alhawi2018, modi:2020} combined network monitoring with ML techniques for ransomware detection. In NetConverse~\cite{Alhawi2018}, Alhawi et al. built a DT classifier using protocol type, IP addresses, number of packets and bytes, and duration features of the network traffic to detect ransomware. Modi et al.~\cite{modi:2020} aimed to detect ransomware in encrypted web traffic by utilizing 28 features including connection features (e.g., flow, payload, and packet features), SSL features (e.g., ratios of SSL flows, SSL-TLS, etc.), and certificate features (e.g., certificate validity, age, etc.) to build RF, SVM, and logistic regression classifiers. 

In terms of the network-based traffic monitoring schemes, Cusack et al.~\cite{Cusack:2018} proposed a solution based on networking hardware, namely Programmable Forwarding Engines to monitor the network traffic between a ransomware infected computer and the C\&C server. During the monitoring phase, they extract standard deviation of packet lengths and number of bytes in inflows and outflows, mean burst length of inflows, minimal interarrival time of outflows, and the ratio of outflow to inflow packets, and build a detection system using a RF classifier.

\vspace{0.2em}
\textbf{\emph{Via API Call Behavior:}}
One of the main behavioral features obtained from dynamic analysis of ransomware is API calls. In this context, the works~\cite{SVMBased, flowgraph, Baeetal, zeroDayAware, avoidingDigital, lstmbased, aliSaleh:2019,AHMED2020102753,Zhouetal:2020} used API calls as features to build ML classifiers to detect ransomware in PCs/workstations. Some of the studies used API calls as features and built SVM classifiers~\cite{SVMBased}, Long-Short Term Memory (LSTM) classifiers~\cite{lstmbased}, Recurrent Neural Network (RNN) classifiers~\cite{recurrentNeural}, and Restricted Boltzmann Machine classifiers~\cite{avoidingDigital}. N-grams of API calls were also used by researchers to build SVM classifiers~\cite{zeroDayAware} and various ML-based classifiers~\cite{Baeetal}. While Chen et al.~\cite{flowgraph} generated API call flow graphs (CFG) and trained different classifiers, Zhou et al.~\cite{Zhouetal:2020} built SVM classifiers using Pearson correlation values of API calls belonging to different API groups.

In addition to the reviewed studies building various classifiers using API calls, some researchers focused more on finding the most significant API call features. Ahmed et al.~\cite{AHMED2020102753} proposed a new filtering method in the feature selection process to find the most appropriate API call n-grams for ransomware detection. They tested the performance of various ML classifiers. Al-Rimy et al.~\cite{aliSaleh:2019}, focused on choosing the most significant API call features and the best classifier combination in an ensemble of classifiers for ransomware detection.

\vspace{0.2em}
\textbf{\emph{Via a Set of Behavioral Features:}}
Some of the studies used a set of behavioral features to build ML classifiers to detect ransomware in PCs/workstations. In this regard, a Bayesian Belief Network (BBN) classifier by Goyal et al.~\cite{Goyal:2020}, an LSTM classifier by Roy and Chen~\cite{DeepRan}, and multiple ML classifiers by Homayoun et al.~\cite{KnowAbnormal.} and Chen et al.~\cite{patternExtraction} were built for ransomware detection. The sets of features to build the classifiers include sequences of events from host logs in Chen et al.~\cite{patternExtraction}, registry changes, file system activity, and DLLs in Homayoun et al.~\cite{KnowAbnormal.}, and ten features including generation rate of encrypted files, file write operations, CPU utilization, deletion of shadow copies, registry changes, file renaming, file size increases, etc. in Goyal et al.~\cite{Goyal:2020}.





\vspace{0.4em}
\noindent\textbf{\emph{Via Both Structural and Behavioral Features:}}
Instead of using only structural or behavioral features, some of the researchers employed features from both groups for ransomware detection. Artificial Neural Networks (ANNs) and SVM classifiers by Abukar et al.~\cite{HSRAutomated}, Markov model and RF classifier by Hwang et al.~\cite{twoStageML}, Naive Bayes and DT classifiers by Zuhair et al.~\cite{zuhairetal}, SVM classifier by Maigida et al.~\cite{migidaetal}, logistic regression classifier by Sgandurra et al.~\cite{eldeRan}, and various ML classifiers by Hasan and Rahman~\cite{ransHaunt}, Egunjobi et al.~\cite{Egunjobietal}, Abbasi et al.~\cite{Abbasietal}, and Ashraf et al.~\cite{Ashrafetal} were built for ransomware detection. While strings are the mostly employed structural feature for the aforementioned studies, API calls, file and directory operations, registry keys, processed and dropped file extensions are the most frequently used behavioral features utilized by these studies to build ML classifiers. Some of the studies employed specific techniques to select the best features for the classifiers. In this regard, Abbasi et al.~\cite{Abbasietal} used Mutual Information (MI) and Particle Swarm Optimization, Ashraf et al.~\cite{Ashrafetal} utilized MI, Principal Component Analysis (PCA), and n-gram techniques, and Maigida et al.~\cite{migidaetal} incorporated Grey Wolf optimization algorithms.

\vspace{0.4em}
\textbf{\emph{Hybrid Detection}} \hfill

In addition to the studies employing one of the aforementioned detection techniques, a few studies exist in the literature that used a set of those approaches. 

Mehnaz et al. proposed RWGuard~\cite{RWGuard}, which employs decoy files monitoring, ML-based process monitoring, file change monitoring, crypto API function hooking, and file classification to detect ransomware. Decoy files are used to detect ransomware-like processes. Process monitoring module trains a number of ML classifiers using number of read, open, create, write, and close I/O requests, and number of temporary files created. File change monitoring module compares the similarity, entropy, file type and sizes before and after the changes in the monitored files. Lastly crypto API function hooking module tries to obtain the encryption keys of processes via hooking techniques. Jethva et al.~\cite{Jethvaetal} proposed a two-layer ransomware detection system that combines ML-based and rule-based techniques. In the first layer, a ML classifier (e.g., SVM, RF, or logistic regression) tries to detect ransomware using API calls, registry key operations, DLLs, enumerated directories, strings, and other features. The rule-based system in the second layer monitors the changes in the file signatures and entropy to detect ransomware.

\vspace{0.4em}
\noindent\textbf{Overview of Ransomware Detection Research for PCs/Workstations: } 

\begin{figure*}[!t]
\vspace{-1.25em}
\centering
\subfigure[Detection Techniques]{%
\includegraphics[scale= 0.24]{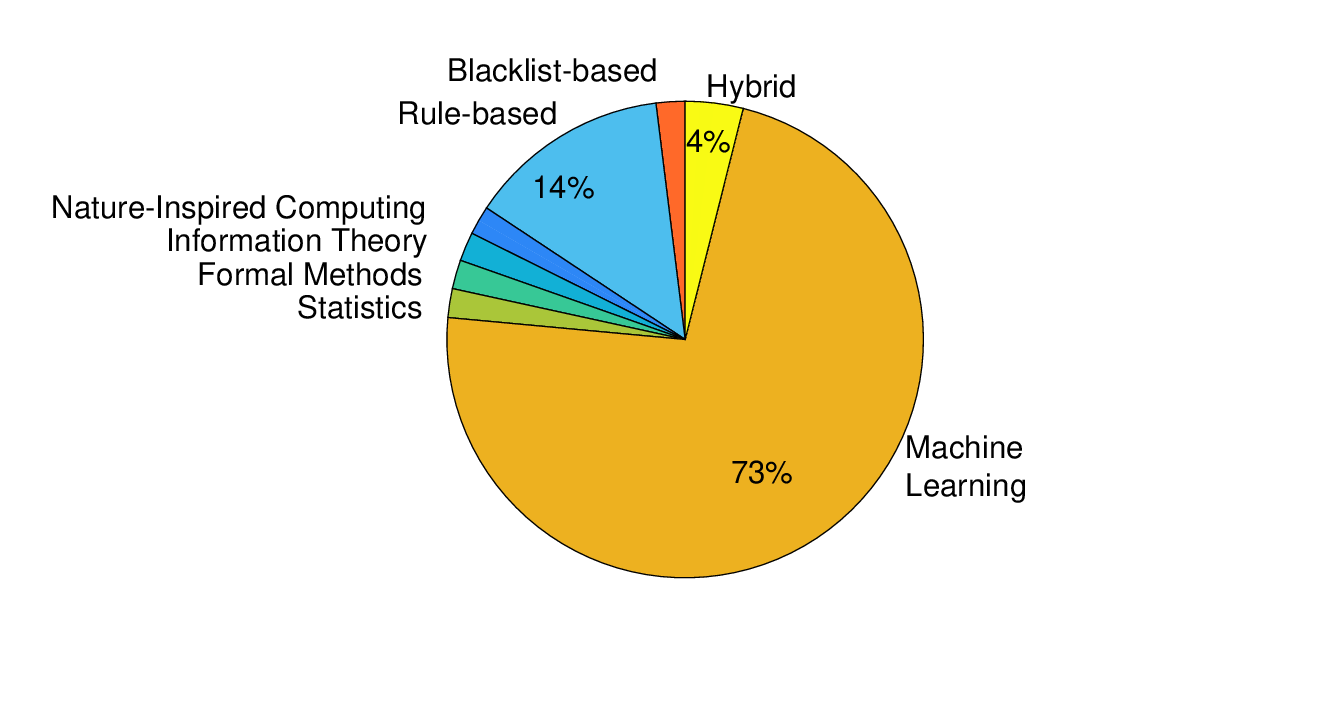}}
\subfigure[Detection Features]{%
\includegraphics[scale= 0.24]{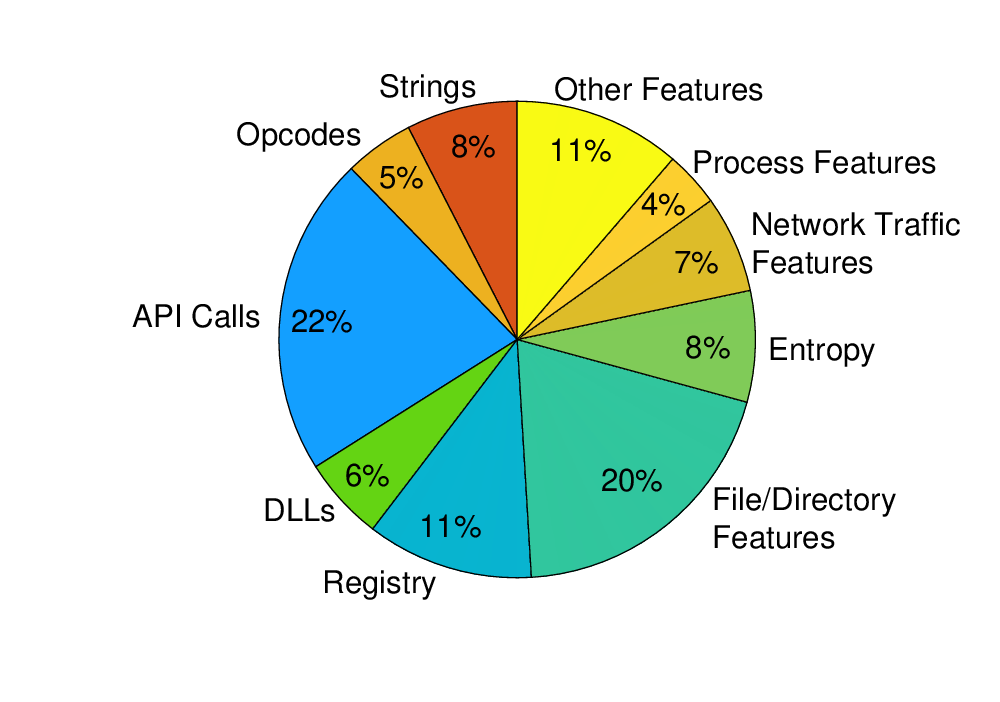}}
\subfigure[Evaluation Dataset]{%
\includegraphics[scale= 0.24]{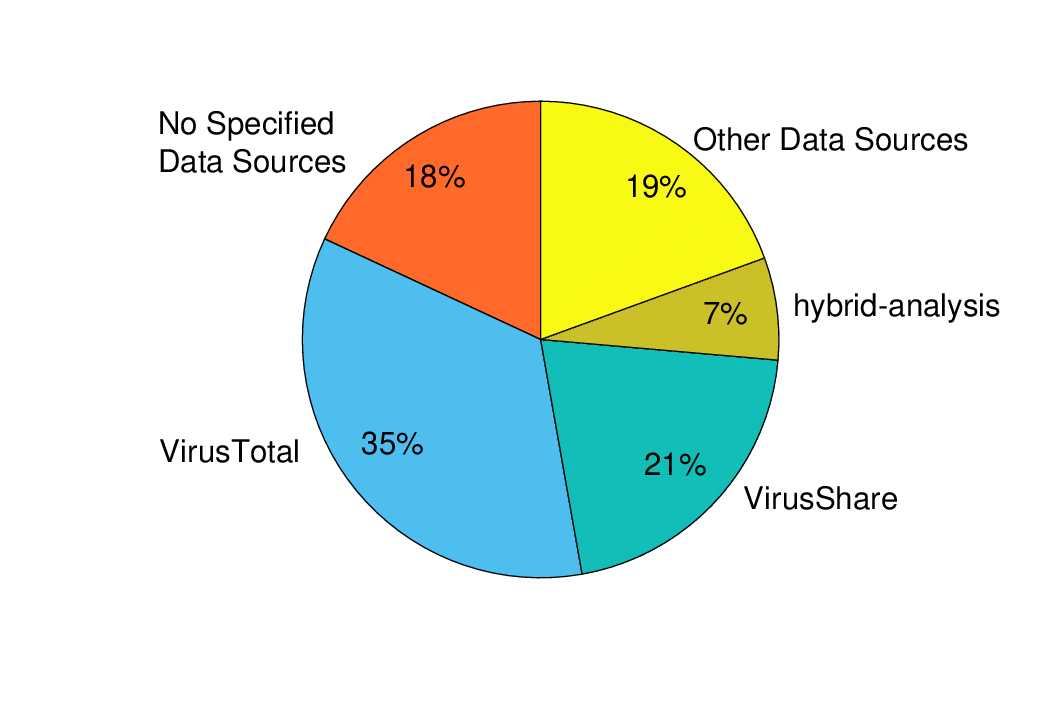}}
\vspace{-1.35em}
\caption{Distribution of detection techniques, detection features, and evaluation datasets employed by the ransomware defense solutions for PCs/workstations.}
\label{fig:pc-tax-graph}
\end{figure*}

The summary of ransomware detection systems for PCs/workstations is given in Table~\ref{tab:pc-workstation-detection}. The table outlines the studies with respect to their techniques, used features, datasets (i.e., data source, ransomware families and corresponding number of ransomware samples, and benign samples), and detection accuracies (i.e., True Positive Rate (TPR) and False Positive Rate (FPR) in \%). Figure~\ref{fig:pc-tax-graph} shows the distribution of techniques, features, and evaluation datasets employed by the studies.

\begin{table}
    \caption{Summary of Ransomware Detection Systems for PCs/workstations.}\label{tab:pc-workstation-detection}
    \centering
    \begin{adjustbox}{width=\textwidth,totalheight=\textheight,keepaspectratio}
     \begin{tabular}{|p{0.8cm}|p{3.4cm}|p{7.3cm}|p{5.1cm}|c|c|c|c|c|}
    \hline
        \multirow{2}{*}{\textbf{Work}} &
  \multirow{2}{*}{\textbf{Detection Technique}} &
  \multirow{2}{*}{\textbf{Features Used}}&
  \multicolumn{4}{c|}{\textbf{Dataset}} &
  \multicolumn{2}{c|}{\textbf{Accuracy Reported}}  \tabularnewline

\cline{4-9} 
    &  & & \textbf{Data Sources Used} & \textbf{\# of Families} &
  \textbf{\# of Malicious} &
  \textbf{\# of Benign}&
  \textbf{TPR} &
  \textbf{FPR}  \tabularnewline
\hline 

\cite{akbanov:2019}&Blacklist-based&Domain names, IP addresses&N/A&1&N/A&N/A&N/A&N/A\\\hline       

\cite{medhatetal}&Rule-based&API calls, strings, file extensions&VirusTotal, hybrid-analysis, MalShare&45&793&878&98.3&8.4  \\\hline

\cite{cryptoDrop}&Rule-based&File type changes and funneling, similarity and entropy of original and modified files, file deletion&VirusTotal&14&492&30&100&1 \\\hline

\cite{EnginKirda}&Rule-based&File system access patterns, I/O data buffer entropy, structural similarity of screenshots&VirusTotal, Anubis, Malwr&15&2,201&49&96.3&0 \\\hline

\cite{redemption}&Rule-based&Entropy ratio of data blocks, file content overwrite, delete operation, directory traversals, conversions to a specific file type, access frequency& Malwareblacklist&29&1,181&230GB&100&0.8 \\\hline

\cite{redFish}&Rule-Based&number of files deleted, time interval between deletion events, average R/W speed&hybrid-analysis, malware-traffic-analysis&19&54&30&100&$\simeq0$ \\\hline

\cite{minetal}&Rule-Based&Intensity, similarity, and entropy of write operations&N/A&N/A&N/A&N/A&N/A&N/A\\\hline

\cite{Cabaj:2016}&Rule-based&HTTP POST message content size&N/A&2&N/A&N/A&97&4.5 \\\hline

\cite{DaD}&Statistics-based&buffer content, size, and offset, file name, process id and name, thread id&VirusShare, MalekalDB&20&798&N/A&99.37&0.41\\\hline

\cite{leeetal2}&Information Theory&Entropy of files&N/A&0&100&100&100&N/A\\\hline

\cite{databasedetect}&Formal Methods-based&Database query sequences&N/A&N/A&N/A&N/A&100&0 \\\hline

\cite{lu:2020}&Nature Inspired Computing-based&API call n-grams&N/A&N/A&2000&1000&96&N/A \\\hline

\cite{Xiao:2018}&ML-Structural Features&Opcode n-grams&VirusTotal&8&1787&N/A&99.8&N/A \\\hline

\cite{ZHANG2020708}&ML-Structural Features&Opcode n-grams& VirusTotal&17&302&N/A&97&N/A \\\hline

\cite{salehetal}&ML-Structural Features&Opcodes&hybrid-analysis, public rep. \cite{github3}&N/A&17&19&73&N/A \\\hline

\cite{SVMbaldwin}&ML-Structural Features&Opcodes&VirusTotal&5&5&1&97.1&0.3 \\\hline

\cite{Martinelli2018PhylogeneticAF}&ML-Structural Features&API calls&VirusTotal&3&91&100&88.5&16.9 \\\hline

\cite{poudyaletal}&ML-Structural Features&Opcodes, DLLs&VirusTotal, VirusShare, public rep.~\cite{github3}&12&178&178&97&N/A\\\hline

\cite{Taylor2017SensorbasedRD}&ML-Hardware Behavior&Sensor readings&Custom&1&1&&95&  \\\hline

\cite{SSD_Insider}&ML-Hardware Behavior&Statistical overwrite features&VirusTotal, public rep. \cite{github3}&8&12&10&100&$\simeq0$ \\\hline
  
\cite{RansomBlocker}&ML-Hardware Behavior&Write operations &N/A&N/A&N/A&N/A&100&0 \\\hline

\cite{COHEN2018158}&ML-Hardware Behavior&DLL, processes, mutexes, services, handles, kernel modules and callbacks&N/A&5&100&100&99&8 \\\hline

\cite{shieldFS}&ML-File System Behavior&Number of files accessed, read, written, renamed or moved, entropy of write operations, folder listing operations&VirusTotal, Custom~\cite{shieldFS}&18&688&2245&97.7&0.038 \\\hline

\cite{modi:2020}&ML-Network Traffic Behavior&Connection, SSL, and certificate features&VirusTotal&20&N/A&30&99&0 \\\hline

\cite{Alhawi2018}&ML-Network Traffic Behavior&Protocol type, IP addresses, number of packets and bytes, and duration& VirusTotal&9&210&264&95&$\simeq3.5$ \\\hline

\cite{Cusack:2018}&ML-Network Traffic Behavior&Packet lengths and number of bytes in inflows and outflows, burst length of inflows, interarrival time of outflows, ratio of outflow to inflow packets&N/A&N/A&100MB&100MB&87&10 \\\hline

\cite{SVMBased}&ML-API Call Behavior&API calls&hybrid-analysis&N/A&276&312&97.48&1.64  \\\hline

\cite{Zhouetal:2020}&ML-API Call Behavior&Correlation of API call frequencies&hybrid-analysis, VirusShare, Virusign, theZoo&9&1140&241&98.2&N/A \\\hline

\cite{zeroDayAware}&ML-API Call Behavior&API call n-grams&VirusShare&4&38,152&1000&99&2.4 \\\hline

\cite{Baeetal}&ML-API Call Behavior&API call n-grams&VirusTotal&58&1000&300&$\simeq 98$&N/A \\\hline

\cite{AHMED2020102753}&ML-API Call Behavior&API call n-grams&VirusShare, VirusTotal&14&1354&1358&97.4&1.6 \\\hline

\cite{lstmbased}&ML-API Call Behavior&API Calls&Online sources, honeynets&N/A&157&N/A&96,67&N/A \\\hline

\cite{flowgraph}&ML-API Call Behavior&API Call Flow Graphs&VirusShare&4&83&85&$\simeq 98$&1.2 \\\hline

\cite{aliSaleh:2019}&ML-API Call Behavior&API calls&VirusTotal&15&8,152&1000&98&7.1 \\\hline

\cite{avoidingDigital}&ML-API Call Behavior&API calls&VirusTotal, VirusShare&14&1232&1308&94.61&5.38 \\\hline

\cite{recurrentNeural}&ML-API Call Behavior&API calls&N/A&N/A&26300&N/A&93&2 \\\hline

\cite{Goyal:2020}&ML-Set of Behavioral Features&Generation rate of encrypted files, file write operations, CPU usage, deletion of shadow copy, registry changes, file renamings, file size changes, wallpaper changes, network activity&VirusShare, VirusTotal, public rep.~\cite{github3}&5&200&N/A&95&0 \\\hline

\cite{KnowAbnormal.}&ML-Set of Behavioral Features&Registry changes, file operations, DLL events&VirusTotal&3&1624&220&99.4&4 \\\hline

\cite{patternExtraction}&ML-Set of Behavioral Features&API calls, file events, registry keys&N/A&7&7&N/A&99&0 \\\hline

\cite{DeepRan}&ML-Set of Behavioral Features&Event sequences in host logs&N/A&17&929,967&4.820&99.87&0\\\hline

\cite{HSRAutomated}&ML-Both Structural and Behavioral Features&API calls, file and directory operations, and registry paths&VirusShare, VirusTotal&14&1254&1308&98.6&2.6 \\\hline

\cite{ransHaunt}&ML-Both Structural and Behavioral Features&Function length frequency, strings, API calls, registry key operations, file operations&VirusShare&21&360&460&97.1&$\simeq 3$ \\\hline

\cite{twoStageML}&ML-Both Structural and Behavioral Features&API calls, registry key operations, file system and directory operations, file extensions and dropped extensions, strings&VirusShare&N/A&1176&1160&97&$\simeq 4.83$\\\hline

\cite{zuhairetal}&ML-Both Structural and Behavioral Features&10 structural and 14 behavioral features including API calls, registry key operations, directory actions, file names and extensions, entropy, PE header and signature&VirusTotal, VirusShare&14&35,000&500&97&2.4\\\hline

\cite{Egunjobietal}&ML-Both Structural and Behavioral Features&Hash value, file size, DLLs, mutexes, PE info&VirusTotal&N/A&200&200&100&1 \\\hline

\cite{migidaetal}&ML-Both Structural and Behavioral Features&API calls, registry key operations, directory and file system operations, operations per file types, dropped files, strings&N/A&11&582&942&99.7&$\simeq 0.1$ \\\hline

\cite{eldeRan}&ML-Both Structural and Behavioral Features&API calls, registry keys, file and directory operations, dropped files, strings&VirusShare&11&582&942&96.3&1.6 \\\hline

\cite{Abbasietal}&ML-Both Structural and Behavioral Features&API calls, extensions of processed and dropped files, registry key operations, file and directory operations, strings&VirusTotal, VirusShare&11&582&942&$\simeq 97.34$&N/A\\\hline

\cite{Ashrafetal}&ML-Both Structural and Behavioral Features&PE header features, strings, API calls, registry key operations, file and directory operations, file extenstions, dropped extensions, network domains, DLLs&VirusTotal, VirusShare&N/A&45,000&3000&$\simeq 92$&$\simeq 3$ \\\hline

\cite{RWGuard}&Hybrid Detection&decoy files, I/O request packages, fastIO requests, temporary files created, file similarity, entropy, type and sizes& VirusTotal, OpenMalware, VXVault, Zelster, Malc0de&14&14&261&100&$\simeq0.1$ \\\hline

\cite{Jethvaetal}&Hybrid Detection&API calls, registry key operations, DLLs, enumerated directories, mutex information, strings, packer entropy, file signatures, file entropy&VirusTotal&20&666&103&$\simeq 100$&1.41 \\\hline





\hline
    \end{tabular}
    \end{adjustbox}
    \end{table}

\vspace{0.2em}
\noindent{\textbf{Detection Techniques:}}
Machine Learning-based detection is the most widely used approach for ransomware detection for PCs/workstations. $73\%$ of the studies employed ML-based detection. Among the ML-based works, majority of the studies used behavioral features ($43\%$) that is followed by the studies using structural features ($12\%$), and both behavioral and structural features ($18\%$). The second popular choice of ransomware detection technique has been the rule-based detection which has been utilized by $14\%$ of the studies. In addition to ML-based and rule-based systems, variety of detection techniques from different domains were used by researchers to detect ransomware as shown in Table~\ref{tab:pc-workstation-detection} and Figure~\ref{fig:pc-tax-graph}(a).



\vspace{0.2em}
\noindent{\textbf{Detection Features:}}
API calls and file/directory features are the most popular features used for ransomware detection for PCs/workstations. Since ransomware performs malicious actions on the file system and makes various API calls while doing its actions, file/directory features and API calls are the most widely looked features for ransomware patterns. The rest of the features are also employed by researchers. However, they are not leveraged as frequent as the API calls and file/directory features. It may be due to these features being platform dependent (e.g., DLLs, registry), or easy to obfuscate (e.g., strings, opcodes, network traffic), or having issues with already compressed file types (e.g., entropy). 

\vspace{0.2em}
\noindent{\textbf{Evaluation Datasets:}}
VirusTotal is the most popular data source for ransomware detection systems for PCs/workstations. It is followed by VirusShare, hybridanalysis.com, and the others. We can see that the majority of the studies employed samples from several ransomware families (the average of number of families used in the datasets is $\simeq10$). As outlined in Table~\ref{tab:pc-workstation-detection}, many studies used more than $1000$ ransomware samples in their datasets. Considering the number of benign samples in the datasets, we can see that some researchers tried to use balanced datasets while the others chose to evaluate their scheme based on an imbalanced dataset. While the majority of the studies reported the number of ransomware families, some studies did not state it.

\vspace{0.2em}
\noindent{\textbf{Detection Accuracy:}}
The ransomware detection studies for PCs/workstations reported very high detection rates. TPR changes between 73\% and 100\%, while FPR changes between 0 and 16.9\%. Many studies reported perfect TPR (i.e., 100\%) that look over-optimistic. We can see that the number of families used in those studies varies between 8 and 29. If the number of employed ransomware families increases, the detection accuracy of some studies may change. 

\noindent\subsubsection{Ransomware Detection for Mobile Devices}\hfill

In this subsection, we categorize and give an overview of ransomware detection systems for mobile devices. Considering the existing works, we can see that rule-based, formal methods-based, machine learning-based, and hybrid detection techniques were employed by researchers. As Android is the most popular target of mobile ransomware as explained in Section~\ref{sec:taxonomy-ransomware}, the detection systems summarized in this subsection are for Android platforms.

\textbf{\emph{Rule-Based Detection}}\hfill

Three rule-based mobile ransomware detection systems were proposed by researchers that use threshold values for detection. RanDroid~\cite{ranDroid} extracts images and strings from applications and calculates their similarity to the images and strings of ransomware samples. Based on the threshold values, it detects mobile ransomware. In the detection system of Song et al.~\cite{Song_Kim_Lee_2016}, modification and deletion events are monitored in a predetermined directory. In case of such events, the proposed system checks if CPU, memory, and I/O usage are above a threshold, and detects ransomware. The last study in this respect is RansomProber proposed by Chen et al.~\cite{chen:Android}. It monitors predefined directories to detect significant changes in entropy. If such a case is detected, then RansomProber tries to understand whether the encryption operation is benign or malicious by trying to match the application performing encryption with the application running in the foreground. Since some applications may look benign but act as ransomware, RansomProber tries detect such applications by checking for user interface elements (i.e., buttons, file list elements, hint text) on the application that benign encryption applications usually display. 




\textbf{\emph{Formal Methods-Based Detection}}\hfill

Formal methods to detect mobile ransomware were employed by two studies in the literature. The defense solution proposed in \cite{Mercaldo:2016} and its extended version in \cite{TalosNM} leveraged Calculus of Communicating Systems (CCS) formal model to detect mobile ransomware. The solutions firstly convert bytecode of applications to CCS model by transforming every instruction in the bytecode into a CCS process. Temporal logic properties of ransomware behavior in CCS model are described. The detection systems perform formal verification using the described temporal logic properties to detect ransomware.

\textbf{\emph{Machine Learning-Based Detection}} \hfill

\vspace{0.2em}
\noindent\textbf{\emph{Via Structural Features:}}
In terms of the ML-based ransomware detection systems for mobile devices using structural features, researchers used API packages~\cite{Alsoghyer_2019,R-PackDroid}, classes, and methods~\cite{scalas:2019}, permissions~\cite{permissionsAndroid}, opcodes in native instruction formats~\cite{nativeMobile}, grey-scale images of mobile application source codes~\cite{Karimietal}, and structural entropy of mobile applications~\cite{stractrualEntropy} to build and evaluate various ML classifiers.

Some researchers aimed to offload the mobile ransomware detection tasks to cloud to save from the resources of mobile devices. In this regard, RanDetector proposed by Alzahrani et al.\cite{alzahranietal} extracts permissions, intents, and cryptography-related API packages in the server-side and use them to train various ML classifiers for ransomware detection. Similarly, the detection system of Faris et al.~\cite{Farisetal} extracts API packages and permissions of mobile applications and uses Salp Swarm Algorithm to select the best features, and utilize Kernel Extreme Learning Machine classifier to detect mobile ransomware.

\vspace{0.2em}

\noindent\textbf{\emph{Via Hardware Behavior:}}
Power usage behavior of mobile applications was used by Azmoodeh et al.~\cite{energyFootPrint} to detect ransomware. They used PowerTutor application to collect power consumption of both benign and ransomware applications at regular intervals, and analyzed the performance of a number of ML classifiers on the collected data. 

\vspace{0.2em}
\noindent\textbf{\emph{Via Both Structural and Behavioral Features:}}

A few studies in the literature aimed to benefit from both static and dynamic analysis of mobile ransomware samples and use the obtained features to build ML models. Ferrante et al.~\cite{ferrante} proposed a mobile ransomware detection system that extracts opcode frequencies via static analysis and obtains CPU, memory, network usage, and
system call statistics via dynamic analysis. In total 87 features were used to train and evaluate various ML classifiers. In DNA-Droid~\cite{DNADROID}, a two-layered detection framework was proposed. The first layer of DNA-Droid consists of a ML classifier that determines the maliciousness score of a sample using the structural features of images, strings, API packages, and permissions. If the sample is determined to be suspicious by the first layer, then the second layer analyzes its API calls during runtime and uses ML classifiers to detect ransomware.

\textbf{\emph{Hybrid Detection}} \hfill

In addition to the studies employing only one of the aforementioned detection techniques, a few studies exist in the literature that used a set of those approaches. In this regard, HelDroid proposed by Andronio et al.~\cite{HelDroid} uses an NLP classifier to detect threatening text of ransomware, employs taint analysis to detect execution flows that signify a ransomware-related encryption operation, and utilizes heuristics with permissions and function calls to detect malicious locking behavior. As another hybrid detection system, GreatEatlon was proposed by Zheng et al.~\cite{greatEeatlon} which aims to improve HelDroid by adding new capabilities to its threatening text, encryption, and locking detectors. GreatEatlon firstly uses an ensemble of ML classifiers using numerous features obtained via static analysis to detect suspicious mobile application packages. Following that, it adds detection of device administration API misuse, reflection misuse, and conditional execution flow controls to detectors of HelDroid to detect mobile ransomware.

\noindent\textbf{Overview of Ransomware Detection Research for Mobile Devices} 

\begin{figure}[!h]
\vspace{-1.25em}
\centering
\subfigure[Detection Techniques]{%
\includegraphics[scale= 0.25]{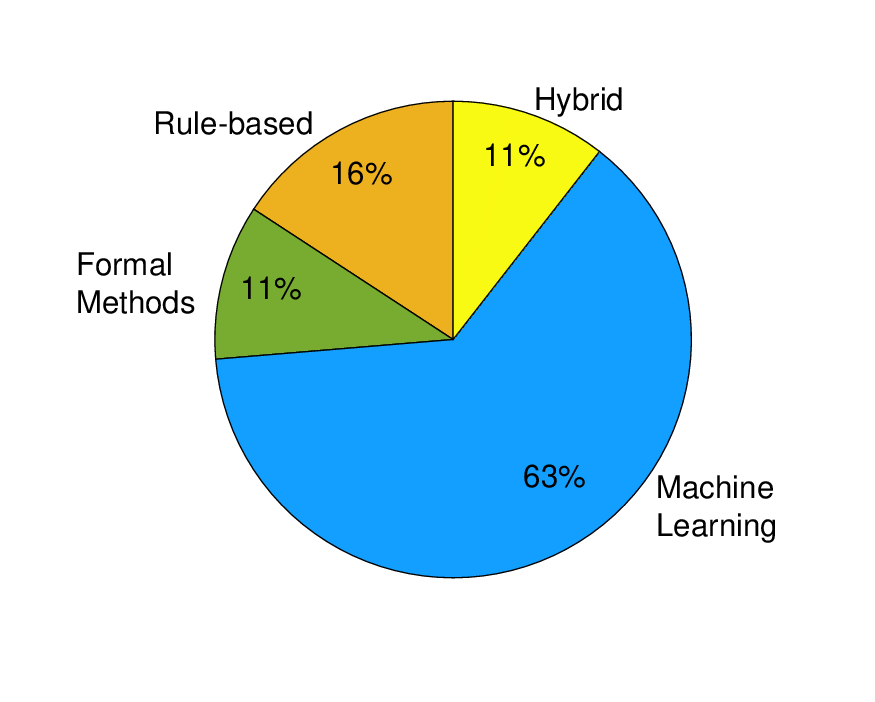}}
\subfigure[Detection Features]{%
\includegraphics[scale= 0.25]{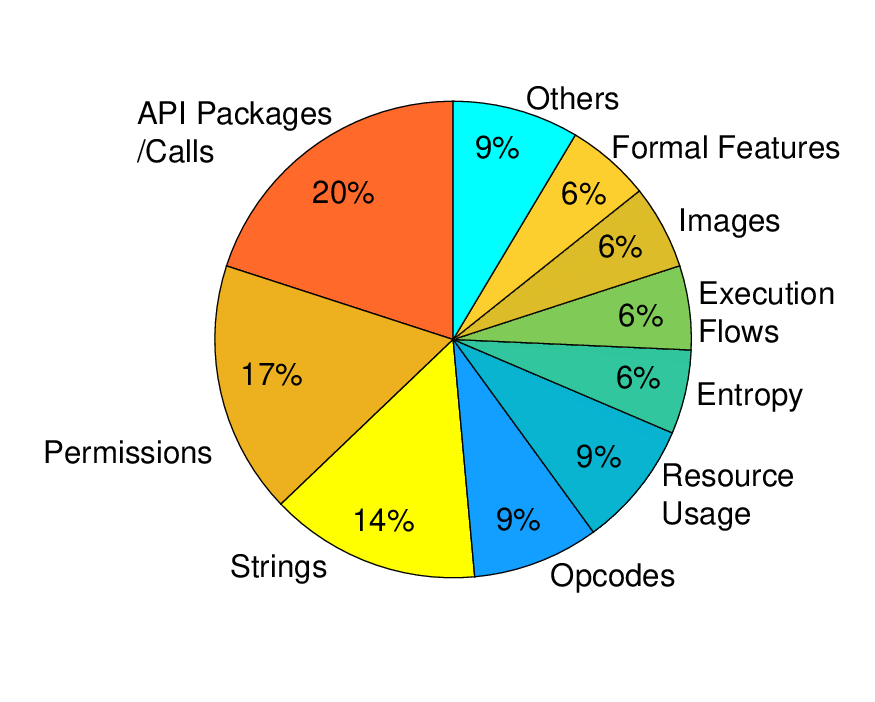}}
\subfigure[Evaluation Dataset]{%
\includegraphics[scale= 0.25]{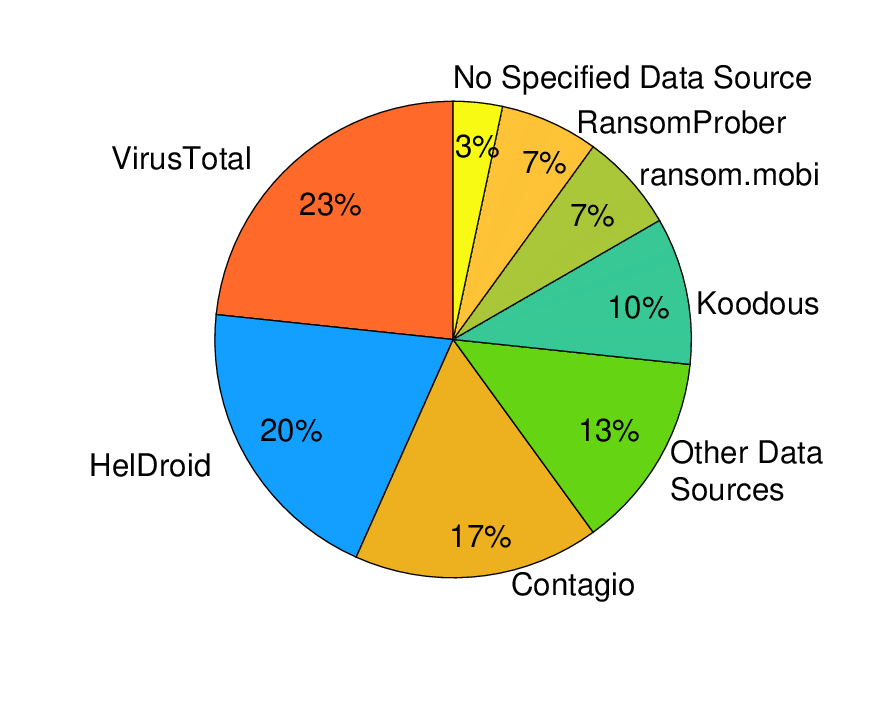}}
\vspace{-1.25em}
\caption{Distribution of detection techniques, detection features, and evaluation datasets employed by the ransomware defense solutions for mobile devices.}
\label{fig:mobile-tax-graph}
\end{figure}

The summary of ransomware detection systems for mobile devices is given in Table~\ref{tab:mobile-devices-detection}. The table outlines the studies with respect to their techniques, used features, datasets (i.e., data source, ransomware families and corresponding number of ransomware samples, and benign samples), and detection accuracies (i.e., TPR and FPR in \%). Figure~\ref{fig:mobile-tax-graph} shows the distribution of techniques, features, and evaluation datasets employed by the studies.

\vspace{0.2em}
\noindent{\textbf{Detection Techniques and Features:}}
As shown in Figure~\ref{fig:mobile-tax-graph}(a), machine learning is the most widely used technique for ransomware detection in mobile devices. Over $60\%$ of mobile ransomware detection systems reviewed in this work employ ML. Considering the utilized features, majority of the studies used structural features that are obtained via static analysis for building ML models. This may be due to the resource limitations of mobile devices which may not be suitable for real-time behavioral analysis of the applications. Rule-based, formal methods-based, and hybrid detection are rest of the techniques incorporated in mobile ransomware detection. 

In terms of the features, API packages/calls is the most popular feature for mobile ransomware detection as Figure~\ref{fig:mobile-tax-graph}(b) shows. API packages/calls, permissions, and strings constitute the $51\%$ of the used features in mobile ransomware detection which shows that one out of every two studies employ either of these features. Considering the features shown in Figure~\ref{fig:mobile-tax-graph}(b), we can see that most of the features are structural features that are obtained via static analysis of application packages.



\begin{table}
    \caption{{Summary of Ransomware Detection Systems for Mobile Devices}}\label{tab:mobile-devices-detection}
    \centering
    \begin{adjustbox}{width=\textwidth,totalheight=\textheight,keepaspectratio}
     \begin{tabular}{|p{0.8cm}|p{3.4cm}|p{7.3cm}|p{5.1cm}|c|c|c|c|c|}    \hline
        \multirow{2}{*}{\textbf{Work}} &
  \multirow{2}{*}{\textbf{Detection Technique}} &
  \multirow{2}{*}{\textbf{Features Used}}&
  \multicolumn{4}{c|}{\textbf{Dataset}} &
  \multicolumn{2}{c|}{\textbf{Accuracy Reported}}  \tabularnewline

\cline{4-9} 
    &  & & \textbf{Data Sources Used} & \textbf{\# of Families} &
  \textbf{\# of Malicious} &
  \textbf{\# of Benign}&
  \textbf{TPR} &
  \textbf{FPR}  \tabularnewline
\hline 

\cite{chen:Android}&Rule-based&Entropy, user interface elements (buttons, file list, hint text)&HelDroid, VirusTotal&4&83&85&97.6&1.2 \\\hline

\cite{Song_Kim_Lee_2016}&Rule-based&File modification and deletion events, CPU, memory and I/O usage&Self-developed&1&1&N/A&N/A&N/A  \\\hline

\cite{ranDroid}&Rule-based&Strings, images&N/A&N/A&100&200&91&- \\\hline

\cite{Mercaldo:2016}&Formal Methods-based&Calculus of Communicating Systems model of application bytecodes&ransom.mobi, Contagio&N/A&1277&600&99.5&0 \\\hline

\cite{TalosNM}&Formal Methods-based&Calculus of Communicating Systems model of application bytecodes&ransom.mobi, Contagio&N/A&1360&1500&98&0.11 \\\hline

\cite{Alsoghyer_2019}&ML-Structural Features&API packages&HelDroid, RansomProber, VirusTotal, Koodous&N/A&500&500&94&$\simeq 3$ \\\hline

\cite{R-PackDroid}&ML-Structural Features&API packages&HelDroid, VirusTotal&N/A&2,047&4,098&97&1 \\\hline

\cite{scalas:2019}&ML-Structural Features&API packages, classes, and methods&VirusTotal, HelDroid&11&3017&N/A&97&1 \\\hline

\cite{permissionsAndroid}&ML-Structural Features&Permissions&HelDroid, RansomProber, VirusTotal, Koodous&N/A&500&500&96.9&3.1 \\\hline

\cite{alzahranietal}&ML-Structural Features&API packages, permissions, intents&Khoron, Contagio&10&259&200&96&1.64 \\\hline

\cite{Farisetal}&ML-Structural Features&API packages, permissions&HelDroid, RansomProber,
VirusTotal, Koodous&N/A&500&500&98&0.2 \\\hline

\cite{stractrualEntropy}&ML-Structural Features&Structural entropy&VirusTotal&N/A&2052&10,000&83&19 \\\hline

\cite{Karimietal}&ML-Structural Features&Opcode sequences&Andrototal&3&250&30&97.5&N/A \\\hline

\cite{nativeMobile}&ML-Structural Features&Native instruction opcodes&Public rep.~\cite{nativeDataset}&6&2148&N/A&99.8&0 \\\hline

\cite{energyFootPrint}&ML-Hardware Behavior&Power consumption of applications&VirusTotal& N/A&6 &12&95.65 &N/A \\\hline

\cite{ferrante}&ML-Both Structural and Behavioral Features&Opcode frequencies, CPU, network, memory usage, system calls&HelDroid&N/A&672&2,386&100&$\simeq 4$ \\\hline

\cite{DNADROID}&ML-Both Structural and Behavioral Features&Images, strings, API packages, permissions, API calls&HelDroid, Contagio, VirusTotal, Koodous&8&1928&2500&97.5&$\simeq 0.5$  \\\hline

\cite{HelDroid}&Hybrid&Strings, execution flows, permission and function call heuristics&HelDroid&N/A &207&14&$\simeq 100$& N/A\\\hline

\cite{greatEeatlon}&Hybrid&Strings, execution flows, permission and function call heuristics, and numerous features&Contagio, VirusTotal&N/A&75&N/A&99&$\simeq 0$ \\\hline



\hline
    \end{tabular}
    \end{adjustbox}
    \end{table}

\vspace{0.2em}
\noindent{\textbf{Evaluation Datasets:}}
The most popular data source for ransomware detection systems for mobile devices are VirusTotal and the dataset of HelDroid~\cite{HelDroid}. These data sources are followed by Contagio, Koodus, and other datasets. We can see that the majority of the studies formed their datasets using multiple data sources. Unlike the case in PCs/workstations, most of the studies for mobile ransomware detection did not report the number of ransomware families in their datasets. In terms of the studies that report, we see at most 10 families were used by the studies. Considering the number of malicious and benign samples, most of the datasets are imbalanced datasets which can better represent the rate of benign and malicious mobile applications in the wild.

\vspace{0.2em}
\noindent{\textbf{Detection Accuracy:}}
The ransomware detection studies for mobile devices reported very high detection rates. TPR changes between 83\% and 100\%, while FPR varies between 0 and 19\%. Only one study reported a perfect TPR (i.e., 100\%), while several studies reported a TPR over 99\%.

\subsubsection{Ransomware Detection for IoT/CPS}\hfill 

Since ransomware detection for IoT/CPS environments is not a well explored field of research, there are only five studies tackling the ransomware detection problem in such environments. Considering the detection studies, all of the studies utilize ML techniques.

\textbf{\emph{Machine Learning-Based Detection}} \hfill

\vspace{0.2em}
\noindent\textbf{\emph{Via Network Traffic Behavior:}}
Considering the ML-based ransomware detection systems for IoT/CPS, there exist two studies. In the first study, Maimó et al.~\cite{clinicalEnv} proposed a ransomware defense system for Integrated Clinical Environments (ICE) of Medical CPS. The proposed system monitors the traffic between the medical CPS devices and the ICE system. By extracting TCP and UDP flow features it detects unseen and known ransomware strains via SVM and Naive Bayes classifiers respectively. In the second study, Wani and Revathi proposed IoTSDN-RAN~\cite{wanietal.} which aims to monitor the network traffic using the SDN controller, and extracts packet size, host IP and destination server address from Constrained Application Protocol (CoAP) headers. The extracted features are used by IoTSDN to train a Naive Bayes classifier with Principal Component Analysis. 

\vspace{0.2em}
\noindent\textbf{\emph{Via a Set of Behavioral Features:}}
Al-Hawawreh and Sitnikova~\cite{deepIoT} proposed a DL-based ransomware detection system for the workstations that are used as host machines of Industrial IoT environments. Their system relies on classical and variational auto-encoders to select the most appropriate features from several behavioral features of API calls, registry keys, file and directory operations. The same authors published another work \cite{IoTStacked} in the same year on the same problem scope that uses only variational auto-encoders. Unlike Al-Hawawreh and Sitnikova, Alrawashdeh and Purdy~\cite{FPGA} focused on hardware-based ransomware detection in IoT and embedded devices. They proposed an FPGA-based hardware implementation of a Deep Belief Network structure that uses several features including file-related features (e.g., extensions, operations, dropped extensions, source files), registry key operations, HTTP methods, and API statistics.

\vspace{0.4em}
\noindent\textbf{Overview of Ransomware Detection Research for IoT/CPS: } 

The summary of ransomware detection systems for IoT/CPS is given in Table~\ref{tab:IoT/CPS-Detection}. The table outlines the studies with respect to their techniques, used features, datasets (i.e., data source, ransomware families and corresponding number of ransomware samples, and benign samples), and detection accuracies (i.e., TPR and FPR in \%).

\vspace{0.2em}
\noindent{\textbf{Detection Techniques and Features:}}
Considering the detection techniques, only machine learning was used by the researchers for the detection of ransomware in IoT/CPS environments. Although all of the studies were proposed for IoT/CPS environments, only IoTSDN-RAN proposed by Wani and Revathi~\cite{wanietal.} truly considers IoT-specific platforms/protocols (i.e., CoAP). In terms of the features, we can see that flow features, API calls, registry keys, file/directory features are extracted by dynamic analysis and used as behavioral features to train ML models.

\begin{table}
    \caption{Summary of Ransomware Detection Systems for IoT/CPS}\label{tab:IoT/CPS-Detection}
    \centering
    \begin{adjustbox}{width=\textwidth,totalheight=\textheight,keepaspectratio}
    \begin{tabular}{|p{0.8cm}|p{3.4cm}|p{7.3cm}|p{5.1cm}|c|c|c|c|c|} 
    \hline
        \multirow{2}{*}{\textbf{Work}} &
  \multirow{2}{*}{\textbf{Detection Technique}} &
  \multirow{2}{*}{\textbf{Features Used}}&
  \multicolumn{4}{c|}{\textbf{Dataset}} &
  \multicolumn{2}{c|}{\textbf{Accuracy Reported}}  \tabularnewline

\cline{4-9} 
    &  & & \textbf{Data Sources Used} & \textbf{\#Families} &
  \textbf{\#Malicious} &
  \textbf{\#Benign}&
  \textbf{TPR} &
  \textbf{FPR}  \tabularnewline
\hline

\cite{clinicalEnv}&ML-Network Traffic Behavior&TCP and UDP flow features&N/A&N/A&26300&N/A&97&2 \\\hline

\cite{wanietal.}&ML-Network Traffic Behavior&Packet size, host and destination IP addresses&N/A&1&78&N/A&98&2.1 \\\hline

\cite{deepIoT}&ML-Set of Behavioral Features&API Calls, Registry keys, file and directory operations&N/A&N/A&582&942&92.53&7.47 \\\hline

\cite{IoTStacked}&ML-Set of Behavioral Features&Registry keys, file/directory operations, API Calls&N/A&N/A&582&942&99.47&13.9 \\\hline

\cite{FPGA}&ML-Set of Behavioral Features&Extensions and dropped extensions, file operations, source files, registry key operations, HTTP methods&VirusTotal&13&158&N/A&91&2.5 \\\hline

\hline
    \end{tabular}
    \end{adjustbox}
    \end{table}

\vspace{0.2em}
\noindent{\textbf{Evaluation Datasets and Detection Accuracy:}}
For the evaluation of the proposed detection systems, majority of the studies did not report any data sources. Similarly, most of the studies did not report the number of ransomware families in their datasets. In terms of detection performance, the ransomware detection studies for IoT/CPS environments reported high detection rates. TPR changes between 91 \% and 99.47\%, while FPR changes between 2\% and 13.9\%. 

\vspace{5pt} 
\subsubsection{\harun{Comparison of Ransomware Detection Techniques Across the Platforms:}} \hfill

\harun{In this subsection, we compare the detection studies in PCs/workstations, mobile devices, and IoT/CPS environments and share our findings with ransomware detection across various platforms.} 

\smallskip

\noindent\textbf{\harun{Comparison of the Detection Techniques:}} \harun{Our analysis disclosed that machine learning is the most admired technique to detect ransomware across all platforms. Specifically, in total 72\% of  defense solutions utilized machine learning to detect ransomware in the system. In addition, given the behavioral variety of ransomware families targeting PC/workstations, researchers utilized seven different techniques to detect ransomware in PC/workstations. On the other hand, researchers utilized only four different techniques to detect ransomware in mobile devices. Since there are only a few works for ransomware detection in IoT/CPS environments, machine learning is the only used technique in this category. Rule-based detection is the second most popular approach to detect ransomware both in PCs/workstations and mobile devices. Our findings show that researchers considered to benefit most from machine learning techniques to detect the patterns of ransomware behavior in the system compared to other techniques. The underlying reason could be related to machine learning models being able to cope better with never before seen samples and capability of generalization compared to other techniques.}

\smallskip

\noindent\textbf{\harun{Comparison of the Used Features:}} \harun{ 
In terms of the used features, our findings show that ransomware detection studies for PCs/workstations and IoT/CPS environments display a different behavior than the studies for mobile devices. Specifically, we see that majority of the machine learning-based ransomware detection systems for PCs/workstations and IoT/CPS environments rely on behavioral features. Whereas, most of the studies for mobile devices utilize structural features. In general, structural features are easier to extract/collect compared to behavioral features as they do not require samples to run and do not necessitate monitoring of the platform. Since mobile devices have considerably less resources than PCs/workstations, structural features could be preferred over behavioral features for mobile devices for this reason. We would like to note that, although ransomware detection studies for IoT/CPS environments use behavioral features similar to PCs/workstations, they accommodate their detection solutions on a resource rich device such as a PC or workstation. Therefore, their posture in this regard does not contradict with the aforementioned analysis.} 

\harun{Considering the actually used features, API-related features such as API calls and API packages in mobile devices were the most used features across all of the platforms. While file/directory features are also very popular for ransomware detection for PCs/workstations, permissions follow API packages in popularity for mobile devices. Although researchers used several other features to detect ransomware, they are not utilized as frequent as the aforementioned features which may be due to those features being platform dependent (e.g., DLLs, registry activities), easy to obfuscate (e.g., strings, opcodes, network traffic), or having issues with already compressed file types (e.g., entropy).} 


\smallskip
\noindent\textbf{\harun{Comparison of the Datasets:}} \harun{The most widely used data source for ransomware detection systems across all platforms is VirusTotal. This finding is not surprising as VirusTotal is a very popular repository for malware research domain and it provides an academic dataset and an API to researchers from academia free of charge. While 76\%  of the ransomware detection systems in PCs/workstations reported the number of families in their dataset, only 36\% of the works in mobile ransomware detection reported number of families in their dataset.  Interestingly, the majority of the ransomware defense solutions for IoT/CPS environments did not disclose any detailed information about their data source. Considering the number of malicious and benign samples in the datasets, we see that although the studies for PCs/workstations constructed both balanced and imbalanced datasets, most of the datasets for ransomware detection in mobile devices are imbalanced which can represent the real world ratio of benign and malicious applications more realistically.} 

\smallskip
\noindent\textbf{\harun{Comparison of the Detection Accuracies:}}
\harun{Generally, all of the reviewed ransomware detection studies reported very high detection rates. Specifically, while TPR fluctuates between 73\% and 100\%, FPR changes between 0 and 19\%. In this regard, many detection systems for PCs/workstations reported 100\% TPR which look over-optimistic. However, we see only one study for mobile devices that reported a perfect TPR. Since the number of families and also the samples used in the evaluation processes play a crucial role in the obtained result, the reported results may probably get more realistic if the proposed schemes are evaluated against a comprehensive dataset of both benign and malicious samples.}

\subsection{Ransomware Recovery Research}
In this subsection, we categorize and summarize existing recovery mechanisms for ransomware with respect to target platforms.

\vspace{0.2em}
\subsubsection{Ransomware Recovery for PCs/Workstations}\hfill 

Ransomware recovery research for PCs/workstations shows that recovery of the destruction performed by ransomware can be achieved in three different ways: \emph{recovery of keys, recovery of files via hardware,} or \emph{recovery of files via cloud backup}. In this subsection we give an overview of the studies under each category respectively.

\vspace{0.2em}
\noindent\emph{\textbf{Recovery of Keys:}} 
Kolodenker et al.~\cite{paybreak2017} proposed PayBreak~\cite{paybreak2017} - a key-escrow mechanism that intends to capture encryption key(s) by hooking the cryptography APIs and decrypt the victim files. Naturally, it is effective only against the ransomware families that call the corresponding cryptography APIs for encryption.

\vspace{0.2em}
\noindent\emph{\textbf{Recovery of Files via Hardware:}} 
The studies presented in this category aim to recover encrypted files of victims by utilizing the characteristics of storage hardware (i.e., SSD). NAND-based SSDs have the ability of out-of-place update feature that preserves a previous version of deleted data until the Garbage Collector (GC) deletes it. This feature was leveraged by ransomware recovery solutions. The works presented in \cite{SSD_Insider, flashGuard,RansomBlocker} create additional backup pages in SSDs to recover the data from ransomware attacks. Alternatively in \cite{minetal}, Min et al. designed an SSD system that performs an automated backup and minimizes the backup space overhead. Their system utilizes a detection component that leverages hardware accelerator to detect the infected pages in the memory.

\vspace{0.2em}
\noindent\emph{\textbf{Recovery of Files via Cloud Backup:}}
Some of the recovery mechanisms in the literature aimed to recover files utilizing cloud environment for backup purposes. Yun et al.~\cite{CLDSafe} proposed a backup system named CLDSafe that is deployed on the cloud. CLDSafe keeps the shadow copies of files to a safe-zone to prevent file loss. It calculates a similarity score between versions of the files to choose which files to back-up. In RockFS~\cite{RockFS}, Matos et al. aimed to make the client-side of the cloud-backed file system more resilient to attacks like ransomware. It allows administrators to recover files via analyzing logs after ransomware incidents. It also aims to secure the cloud access credentials of users that are stored in the client-side via encryption using the secretly shared key.

\vspace{0.4em}
\noindent\subsubsection{Ransomware Recovery for Mobile Devices}\hfill 

Considering the recovery solutions for mobile devices to enable data recovery from ransomware attacks, there exist only two studies. MimosaFTL~\cite{mimosaFTL} was designed as a recovery-based ransomware defense strategy for mobile devices that are equipped with flash memory as external storage. It collects the access behaviors of ransomware samples and applies K-mean clustering to identify the unique access patterns to the Flash Transaction Layer. In \cite{yalewetal} Yalew et al. aimed to recover from ransomware by periodically performing backups to an external storage. 

\subsection{Other Ransomware Defense Research}
Ransomware defense is a very active topic of research. In this subsection we give a brief overview of rest of the defense studies that do not fall under the categorization applied earlier. These studies can be grouped into moving target, access control, and holistic defense categories.

A moving target defense technique was proposed by Lee et al.~\cite{movingTarget} for ransomware protection that changes the file extensions randomly. 

In terms of the access control mechanisms, Genç et al.~\cite{nomoreransom} proposed UShallNotPass that aims to prevent ransomware attack before performing encryption by blocking the access of unauthorized applications to the pseudo-random number generator functions in the operating system. Another ransomware prevention mechanism named Key-SSD~\cite{key-ssd} implemented a disk-level access control to SSD storage units to  prevent the access of unauthorized applications to the SSD.

Considering the holistic defense systems, Keong et al. proposed VoterChoice~\cite{voterChoice} that uses Suricata Intrusion Prevention System to detect malicious activities. Once such an activity is detected, ML-based detection modules that use encryption and registry activities as features detect ransomware. If ransomware is detected, then a client based-honeypot collects activities of the sample to understand the behavior. Jung et al. \cite{jung:2018} proposed a ransomware defense system that consists of monitoring, detection, secure zone file backup, and gray list modules. API calls of applications are monitored by the monitoring module to detect ransomware. If a suspicious process is detected, then the entropy of the modified file is used to determine if the application is ransomware. If a large number of read/write operations are detected, then the secure zone component backs up all the files that are accessed by the application. Shaukat et al.~\cite{ransomwall} proposed a defense system that implements a honey files-based trap-layer and an ML-based detection layer. It uses a set of features such as API calls, registry modifications, deletion of shadow copies, and file system operations to train ML classifiers. It also backs up user files when the trap layer detects ransomware.


\section{OPEN ISSUES}\label{sec:lessons}
Considering the evolution and taxonomy of ransomware, and ransomware defense research for PCs/workstations, mobile devices, and IoT/CPS environments, it is crucial to highlight the open issues in ransomware research.


\vspace{0.4em}

\noindent{\textbf{The Constant Evolution of Ransomware:}} 
Ransomware has been evolving since the appearance of the first ransomware in 1989. It has been changing its target platforms and users, infection methods, encryption techniques, communication mechanisms, destruction behavior, and payment methods. Currently, a ransomware can target various platforms, use numerous infection vectors, utilize dynamically generated domains, TOR network, bitcoin, encrypted communications, employ strong AES and RSA, non-reversably destruct the target platform, steal information, and get paid without easily being traced. However, this is not the end of the story. Ransomware keeps evolving to continue the arms race against defense systems. Here, we enumerate the distinct and modern malicious tactics of emerging ransomware families that future ransomware research can address.

\textbf{Human-Operated Ransomware Attacks.} Unlike auto-spreading ransomware like WannaCry or NotPetya, skilled cybercriminals have started to perform human-operated ransomware campaigns to business organizations. Unlike traditional ransomware which perform infection and malicious actions in an automated manner, these steps of ransomware are performed by human operators in such attacks that have deep knowledge in systems. For this reason, defenders have to combat against attackers in real-time rather than combating against ransomware binaries running autonomously. In addition to traditional ransomware actions performed in these attacks, human operators utilize other malicious payloads, steal data, and spread ransomware~\cite{MicrosoftSecurity_2020}. Human-operated ransomware can pose a new dimension in ransomware defense research.

\textbf{Rootkit Fashion.} Some ransomware families (e.g., Thanos~\cite{Falcone_2020}) started to utilize rootkit techniques to preserve their secrecy~\cite{Rootkit}. Such ransomware can try to hide itself in the target platform to avoid detection and also delay its execution for after some time rather than executing soon~\cite{timeBomb}. Such a behavior can negatively affect the detection accuracy of the existing systems. 

\textbf{Ransomware Living of the Land.} Recently, some ransomware families like Netwalker~\cite{Netwalker} started to utilize the legitimate applications (i.e., Powershell) to carry out their destructive behavior. Such attacks are called as \emph{Ransomware Living of the Land} or \emph{fileless ransomware}~\cite{Netwalker_Fileless}. Since such ransomware execute malicious actions utilizing benign tools of the target platform, they do not leave any footprint in the system, and the detection of such ransomware becomes very tricky~\cite{fileless}.

\textbf{Changing Encryption Tradition.} Traditionally, ransomware strains aimed to encrypt as many files as possible once the system is infected. This behavior generates a distinct I/O pattern in the low-level that helps to differentiate ransomware from benign applications~\cite{cuttingGordian}. However, cybercriminals can change their encryption tradition in a way that they do not aggressively encrypt the victim files and throttle the operation to be undetected. However, it is a question how existing defense systems would response to such evasive actions of ransomware authors.

\textbf{More Exfiltration Attacks.} The main destructive tactic employed by ransomware was holding the victim's data using encryption, or locking the system unless the requested ransom amount is paid. So most of the defense solutions have been developed against such vicious attempts. However, ransomware gangs recently started to steal information to threaten the victim to publish information~\cite{ZDnet-Stealing-Ransomware}. Since stolen data may contain the user's or company's sensitive information, publication of such data may affect the company or victim detrimentally.

\textbf{Leveraging Internal Threats.} Until now, ransomware was infecting the enterprise systems via traditional malware infection methods such as exploit kits, drive-by downloads, brute force attempts, or spam emails. These traditional methods might be ineffective towards infecting well-protected systems of large business organizations. To bypass these systems, cybercriminals have started to bribe insiders like company employees to install ransomware. One such incident was recently detected for Tesla~\cite{tesla2}. For this reason, it is essential to consider the internal threats that can make the infection process much easier for ransomware. Such an insider attacker can try to disable the existing defense systems, or install ransomware to unprotected segments of the network. 

\vspace{0.4em}

\noindent{\textbf{New Ransomware Targets:}} 
To the best of our knowledge, ransomware strains have not been targeting IoT/CPS platforms in the way they have been targeting other platforms. We believe that ransomware attacks to IoT/CPS devices can be much more severe given the ubiquitous nature of such environments. For instance, ransomware can target the implantable or ambulatory medical devices of patients, and threaten to disrupt the services of such devices unless a ransom is paid. ICS that drive the safety critical systems can also be targeted by ransomware. Considering the fact that PLCs and other ICS devices are not updated and used for decades, a ransomware infecting such environments can have catastrophic effects. \ahmet{ In addition, as autonomous vehicles (e.g., cars, drones, trains, ships, etc.) is an active field of research and practice nowadays, future ransomware strains can target such environments, too~\cite{mekdad2021survey}}. In fact, security researchers created a PoC ransomware that targets smart cars recently~\cite{wannadrive}.
We believe that all of these emerging platforms can be a target for future ransomware strains, and more research is needed in both possible ways to perform ransomware on such platforms and the corresponding defense mechanisms.


\vspace{0.4em}

\noindent{\textbf{Success Factors of Ransomware:}}
A very crucial question to ask is why ransomware is successful despite the existing defensive efforts from both industry and academia. Undoubtedly this question can have numerous answers. However, although possibly not complete, we believe that the following factors can be the major driving sources behind the success of ransomware.

\textbf{Delayed Upgrades or Critical Software Patches.} While ransomware most commonly infects the victims via spam e-mails, it can also employ vulnerabilities in the system software or other applications. Although upgrades and patches may aim to repair such vulnerabilities, it is vital not to delay upgrades or security-related software patches to prevent the infection. However, past experiences with notorious SamSam and WannaCry ransomware strains showed that administrators fail to timely apply upgrades or critical software patches. 

\textbf{Security (Un)aware End-Users.} Another crucial factor behind the success of ransomware is regarding the end-users. Although there is a debate as to whether we should expect security awareness from the end-users or not~\cite{Schneier:2016}, we believe that security awareness in end-users can play a crucial role to make the existing defense solutions stronger. Security training of end-users in terms of the infection vectors of ransomware is very vital. 

\textbf{Effect of the Pandemic and Extraordinary Conditions.} As of the time of writing this survey, the pandemic situation of COVID-19 has been affecting all around the world. Unsurprisingly, ransomware authors have been trying to benefit from the pandemic. Many organizations became vulnerable to ransomware by forcing their employers to work remotely. Moreover, there have been ransomware campaigns that target healthcare-related organizations that become vulnerable due to COVID-19. On the other hand, ransomware attacks to other organizations such as schools decreased~\cite{Freed_2020}. We believe that pandemic situations and other extraordinary conditions (e.g., natural disasters, political events, etc.) can be benefited by malware authors to infect more victims.  

\textbf{Willingness to Pay.} As ransomware evolved to target business organizations rather than ordinary end-users, and the proliferation of payment options, the amount of ransom has significantly increased. Adversaries started to get thousands or even millions of dollars as a reward for their attacks to the business organizations. Indeed, one of the major success factor of ransomware is the victim business organizations willing to pay the demanded ransom. Since the obtained rewards are significant, it enables ransomware human resources to hire more skilled attackers. Recently, some ransomware gangs started to combine their forces to hit larger enterprises in the hope of getting more ransom. As several researchers pointed out this issue, we believe that ransomware will continue to be a great threat as long as victims keep on paying to them.  

\vspace{0.4em}
\noindent{\textbf{Comprehensiveness of Defense Solutions:}}  We see that the majority of defense solutions lack of comprehensiveness. In other words, the employed methods in those systems are only effective against specific types of ransomware families. We believe that such defense solutions can have serious practical issues. Ransomware can have a variety of infection vectors, encryption techniques, communication behaviors, and destruction approaches. However the defense solutions that focus on specific parameters (e.g., crypto API calls, traffic traces of specific protocols, IP addresses, registry activities, strings, etc.) can be useless against the ransomware families that employ other techniques. 

\vspace{0.4em}
\noindent{\textbf{Hardware vs. Software-Based Solutions:}} The majority of ransomware defense solutions are software-based. However, if a ransomware can obtain administrator privileges, it can disable such defense mechanisms. For this reason, alternate defense solutions are needed that cannot be easily disabled by such kernel level ransomware. There exist a few hardware-based defense solutions in the literature to detect ransomware. However those solutions are limited to protect the platforms that utilize a specific storage hardware (e.g., SSDs or a specific class of SSDs). We believe that novel defense solutions are needed against kernel level ransomware. 

\vspace{0.4em}
\noindent{\textbf{Adversarial Machine Learning Attacks:}} As analyzed in the previous section, majority of the defense solutions use ML. While the utilization of ML techniques increases the accuracy and enables to effectively detect never-before-seen ransomware samples, recent studies showed that ML-based classifiers are vulnerable attacks that may manipulate either the training data or test data to bypass detection~\cite{adversarial}. Such attacks are called as Adversarial ML attacks, and have been applied not only in the computer vision domain, but also other domains including malware. The adversarial ML attacks in malware domain mostly target ML classifiers that use structural features. Since ransomware detection for both PCs/workstations and mobile devices have several classifiers using structural features, those classifiers can be targets of adversarial ML attacks. Although such attacks and the corresponding defenses have been researched for general malware domain, it is a topic of research if one can directly apply such attacks or employ the proposed defense solutions for ransomware.

\section{Conclusion}
\label{section:conclusion}

In this paper, we provided a comprehensive survey of ransomware and ransomware defense research with respect to PCs/workstations, mobile devices and IoT/CPS environments. We presented a detailed overview on how ransomware evolved in time, thoroughly analyzed the key building blocks of ransomware, proposed a taxonomy of notable ransomware families, and provided an extensive overview of ransomware defense research including analysis, detection and recovery techniques with respect to various platforms. In addition to these, we derived a list of open research problems that need to be addressed in future ransomware research and practice. As ransomware is already prevalent in PCs/workstations, is becoming more prevalent in mobile devices, and has already hit IoT/CPS recently, and will likely grow further in the IoT/CPS domain very quickly, we believe that this paper will play a crucial role in understanding ransomware research with respect to target platforms and motivating further research.

\section*{ACKNOWLEDGEMENTS}
This work is partially supported by the US National Science Foundation Awards: NSF-CAREER-CNS-1453647 and   NSF-1718116. 
The views expressed are those of the authors only, not of the funding agencies.

\bibliographystyle{ACM-Reference-Format}
\bibliography{ransom.bib}

\end{document}